\documentclass[aps,prb,preprint,showpacs,floatfix]{revtex4-1}
\usepackage{amsmath}
\usepackage{amssymb}
\usepackage{graphicx,color}
\usepackage{epstopdf}
\usepackage{ulem}

\makeatletter

\newcommand{\Rmnum}[1]{\expandafter\@slowromancap\romannumeral #1@}
\makeatother

\begin{document}

\title{Spin Logic via Controlled Correlation in Nanomagnet-Dirac Fermion Heterostructures}

\author{Xiaopeng Duan}
\affiliation{Department of Electrical and Computer Engineering, North Carolina State University, Raleigh, NC 27695-7911, USA}

\author{Yuriy G. Semenov}
\affiliation{Department of Electrical and Computer Engineering, North Carolina State University, Raleigh, NC 27695-7911, USA}

\author{Ki Wook Kim}\email{kwk@ncsu.edu}
\affiliation{Department of Electrical and Computer Engineering, North Carolina State University, Raleigh, NC 27695-7911, USA}


\begin{abstract}
A hybrid structure combining the advantages of topological insulator (TI), dielectric ferromagnet
(FM), and graphene is investigated to realize the electrically controlled correlation between electronic and magnetic subsystems for low-power, high-functional applications. Two-dimensional Dirac fermion states provide an ideal environment to facilitate strong coupling through the surface interactions with proximate materials. The unique properties of FM-TI and FM-graphene interfaces make it possible for active "manipulation" and "propagation", respectively, of the information state variable based solely on the spin logic platform through electrical gate biases.  Our theoretical analysis verifies the feasibility of the concept for logic application with both current-driven and current-less interconnect approaches.  The device/circuit characteristics are also examined in realistic conditions, suggesting the desired low-power performance with the estimated energy consumption for COPY/NOT as low as the \textit{attojoule} level.
\end{abstract}

\maketitle

\section{Introduction} \label{sec1}
Efficient electric control of magnetic/spin states has long been desired for low power and highly functional logic/memory devices.~\cite{Chiba2008,Wang2013} In the context of spintronics, the topological insulator (TI) and graphene each represents a unique extreme:~\cite{Pesin2012,Hasan2010,Han2012} spin-manipulating and spin-conserving, respectively.
The TI surface electron states are topologically protected, in which the spin is locked to the momentum. As a result, its surface state is sensitive to magnetic exchange interactions that break the time reversal symmetry.~\cite{Hasan2010} In fact, the anticipated alteration of electronic structures has recently been observed on the Fe or Cr doped $\mathrm{Bi_{2}Se_{3}}$ surfaces both experimentally and by first principle calculations.~\cite{Chen2010,Kou2012,Schlenk2013,Zhang2012d} Thus the inter-dependence between TI surface transport properties and the proximate magnet magnetization is readily predictable.~\cite{Hasan2010,Kong2011,Yokoyama2011}
In comparison, graphene has extremely weak spin-orbit coupling;  the spin and the momentum can be treated two independent quantum numbers and the spin relaxation length reach several microns.~\cite{Han2012}  Yet, the two-dimensional nature of the graphene crystal enables strong surface interaction with a proximate ferromagnet (FM) that can induce electron spin polarization.~\cite{Haugen2008a,Yang2013,Swartz2012} The linear dispersion relation of graphene electrons also indicates that the induced spin polarization can be controlled electrostatically.~\cite{Duan2012}

Hence, the combination of TI, nanomagnet and graphene possesses ideal qualities to meet the two major requirements of logic device design: (i) manipulation of the desired information state variable (i.e., magnetization) by electrostatic control at the TI-FM interface and (ii) robust propagation of information via the (FM induced) spin polarization in the graphene interconnect. As no lattice displacement is involved, no risk of structural instability exists in contract to normal strain based multiferroic materials. Moreover, dynamical control of magnetic susceptibility and consequent logic reconfiguration offered by Dirac fermions~\cite{Duan2012} is difficult to be matched in the metal based spin circuits.

In this paper, such a logic device is put forth and evaluated in detail. Section~\ref{sec2} outlines the overall operating principles following the Bennett clocking scheme. Two approaches for information transfer between adjacent cells are formulated and modeled in Sec.~\ref{sec3}, followed by the discussion on logic circuit  in terms of a 1-bit full adder (Sec.~\ref{sec4}).  Performance issues and the switching reliability at room temperature are also addressed at the end (Sec.~\ref{sec5}).

\section{Spin logic operation principles} \label{sec2}
\begin{figure}
\includegraphics[width=8.5cm, angle=0]{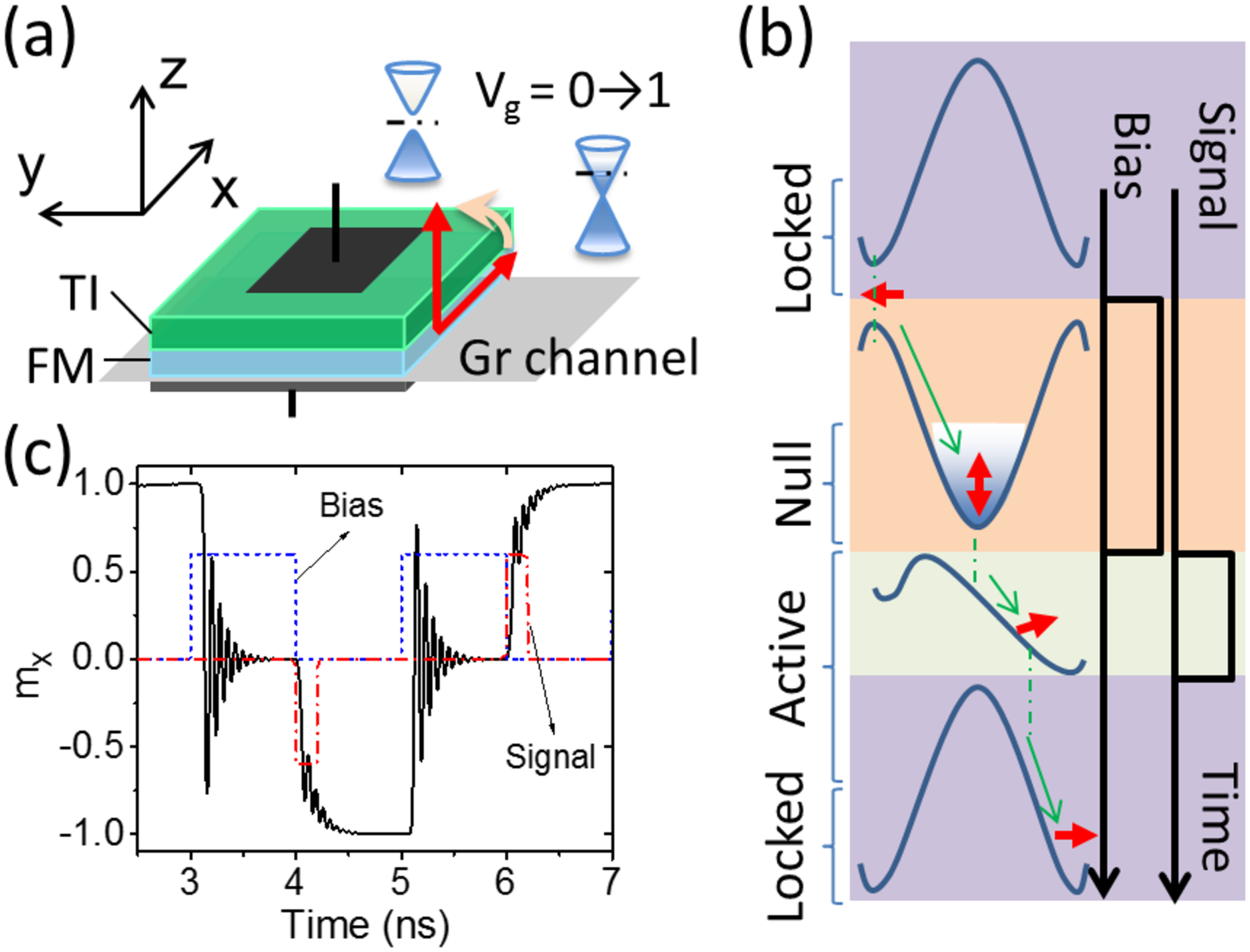}
\caption{(Color online)
(a) The basic component consists of a two-layer structure of topological insulator (TI) and ferromagnet (FM) plus the control gates.  With the gate bias, an effective out-of-plane anisotropy can be induced in the magnetic layer that rotates the magnetization by 90$^{\circ}$ from the in-plane orientation.\cite{Semenov2012}
The graphene (Gr) channel interconnects the elemental cells.  (b) In the Bennett clocking scheme, the bias induced energy minimum constitutes the meta-stable state, where the system resides at the end of the biasing stage (Null). A signal pulse applied subsequently provides an additional effective magnetic field to tilt the free-energy landscape (Active). The arrows indicate  evolution of the magnetization in the ideal conditions. After the relaxation, the magnetization is locked to a stable state along the easy axis. (c) Snapshot of magnetization evolution for $180^{\circ}$ switches with Bennett clocking in the time domain (represented by $m_{x}$). Bias on the TI gate and signal pulses through the channel are indicated by the  dashed lines.}
\label{fig_cell}
\end{figure}

The proposed logic cell consists of a FM-TI stack placed on top of the graphene channel as shown in Fig.~\ref{fig_cell}(a). The top and bottom gate electrodes are separated from the active region by a thin dielectric respectively. The FM is assumed to be insulating or dielectric such as Y$_3$Fe$_5$O$_{12}$ or Fe$_{7}$Se$_{8}$;~\cite{Ji2012} the metallic magnets could cause unintended changes in the TI surface electron density and thus are not desirable.
The interface between the FM and the TI is used to locally control the magnet, while that between the FM and the graphene channel supports the means of interconnect. More specifically, the information is encoded in the magnetization orientation of the magnet, which is then transferred to the electron spin polarization in the graphene layer for dissemination. A functional combination of these two interfaces with the information carrying FM in the middle enables straightforward implementation of Bennett clocking.

The Bennett clocking [Fig.~\ref{fig_cell}(b)] refers to the magnetic switching scheme that uses one clock to put the magnetization in a meta-stable state (the null stage) and another clock to apply the signal that generates a small tilt to determine the final state (the active stage).~\cite{Bennett1982a} If the first clock that overcomes the barrier is applied electrostatically and the critical signal is small in absence of the barrier, this scheme is expected to offer very low energy consumption.
In the present device, the first stage of Bennet clocking is achieved electrically by applying a proper bias at the top gate.  It has recently been demonstrated by a theoretical study that the
correlation interaction in the TI-FM hybrid structure can induce a transition in the easy axis of the magnet between the in-plane and the out-of-plane directions as the TI surface carrier density changes.~\cite{Semenov2012} The resulting 90$^\circ$ rotation of the magnetization to the vertical orientation, once the bias is withdrawn, constitutes the meta-stable state in the Bennett clocking scheme.  Spin polarized electrons in the graphene channel provide the second effective magnetic field that tilts the magnetization slightly toward the desired relaxation direction (i.e., the active stage).  The durations of bias and signal pulses are typically of the order of 1 ns and its fraction, respectively, as shown in Fig.~\ref{fig_cell}(c).  The minimum pulse length depends on material
properties and thermal noise.

In a circuit implementation, the magnetic input signal that determines the final state must be supplied by the preceding cell(s). Unlike metallic magnets that can directly inject polarized electrons into the interconnect medium to output the information, the insulating magnet does not have free carriers.  Instead, the controllable surface exchange interaction at the FM-graphene interface offers an alternatively mechanism to directly induce a spin dependent behavior in graphene that supports transmission of spin information. Specifically, two types of operations are possible; one with and the other without the involvement of electrical current flow. The first scheme relies on the spin dependent carrier transport induced by the exchange barrier at the FM-graphene interface, while the other takes advantage of graphene electron mediated coupling between adjacent magnets. The corresponding physical accounts for each outlined dynamics are detailed in the following.


\section{Information transfer between adjacent cells} \label{sec3}
For an efficient spin logic implementation, it is preferred that the information is transferred in the form of electron spin polarization to avoid the intrinsically inefficient conversion to the electric current. One constraint, however, is the limited distance for reliable signals, which is set by the spin relaxation length. As such, transmission of the information is often accomplished in a cascade, where the state propagates cell by cell along the path. Thus, the issue of information transfer is essentially the interaction between the adjacent cells. At the same time, it is highly desirable if both duplication (COPY) and inversion (NOT) of the upstream spin state can be realized in each of the cascading stage with a relatively simple control and layout arrangement.  These are the underlying principles that motivate the adopted approaches.

\subsection{Via spin polarized electrical currents}
\begin{figure}
\includegraphics[width=7cm, angle=0]{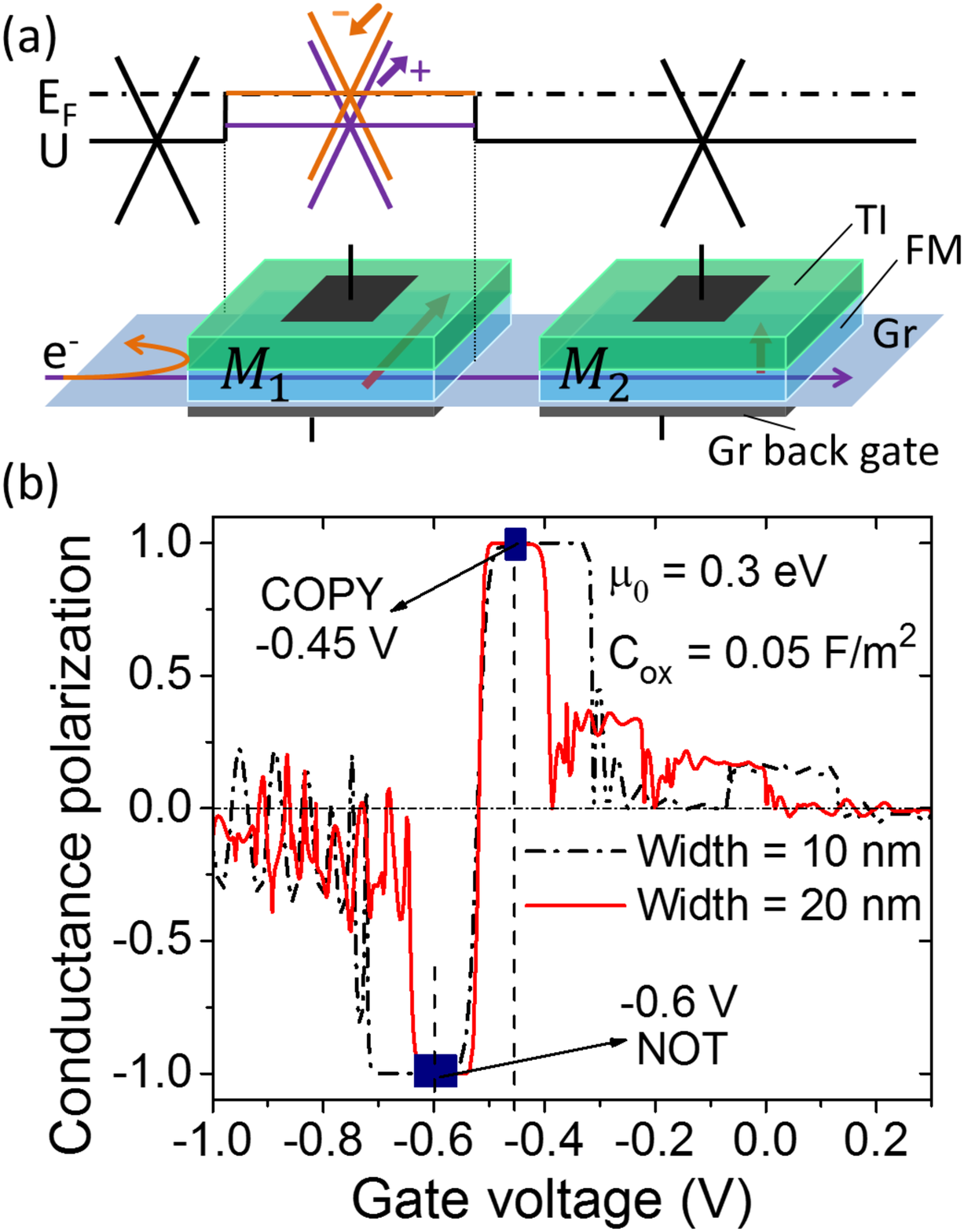}
\caption{(Color online)
(a) COPY/NOT connection of two unit cells. The surface exchange interaction with the magnet induces a spin dependent barrier in graphene.   By controlling electron transmission through the spin split bands (via the back gate bias at $\mathrm{\bf M}_1$), spin polarization of electrons arriving at $\mathrm{\bf M}_2$ can be selected. (b) Calculated conductance polarization in the graphene interconnect as a function of gate voltage. The high polarization windows are marked by the filled rectangles.}
\label{fig_condr}
\end{figure}

The concept of information transfer based on the spin polarized electric current is shown in Fig.~\ref{fig_condr}(a), where the state of magnetization $\mathrm{{\bf M}_2}$ is determined by $\mathrm{{\bf M}_1}$ in the Bennett clocking.  Through the exchange interaction with $\mathrm{{\bf M}_1}$ ($\| \hat{\mathbf{x}}$),
the graphene band structure in the upstream cell lifts the spin degeneracy. Thus, the incoming electrons from the left experience different potential barriers for the spin states parallel and antiparallel to $\mathrm{{\bf M}_1}$, giving rise to a spin dependent conductance.
The transmission probability depends on the quantum state of electrons that induces  Klein tunneling.  The resulting expression, when an electron of energy $E$ in the graphene channel encounters a potential barrier of $U_{g}$, is found to be:\cite{Haugen2008a}
\begin{equation}
T=\frac{(\xi^2-u^2)(1-u^2)}{(\xi^2-u^2)(1-u^2)+u^2(1-\xi)^2\sin^2{(k_{F}L\sqrt{\xi^2-u^2})}}\,,
\end{equation}
where  $\xi=(E-U_{g})/E$ and $u=k_{x}/k$ ($k=\frac{\vert E\vert}{\hbar v_{F}}$; the magnitude of wave vector $\mathbf{k}$). In addition, the transverse wave vector ($k_{y}$) is quantized in a narrow channel and expressed as $k_{y} =(n+1/2)\pi/W$, where $W$ is the channel width and the integer quantum number $n$ is confined within [0, $ \frac{kW}{\pi} - \frac{1}{2} $]. Considering the contribution from multiple energy levels at a finite temperature and the valley degeneracy $g_{v}=2$, the conductance of one spin channel is calculated in the Landauer-B{\"u}ttiker formalism as
\begin{equation}
G(E_{F},U_{g})=g_{v}\frac{e^{2}}{\pi h}\int_{-\infty}^{\infty}dE\sum_{n}T(n,E,U_{g}) \left[ -\frac{\partial f(E,E_{F},T)}{\partial E} \right] ,
\label{eq_LB}
\end{equation}
where $f(E,E_{F},T)$ is the Fermi-Dirac distribution.

The signal strength is determined by the spin polarization of the conductance [Fig.~\ref{fig_condr}(b)], which is defined as the ratio of the difference between the two spin channel conductances over the total conductance.  The quantization step determined by the adopted extreme quantum limit ($=\pi\hbar v_{F}/W$) is approximately $0.2$~eV and $0.1$~eV for the channel width of $10$~nm and $20$~nm, respectively,~\cite{Akhmerov2008} sufficiently large even for room temperature. The conductance approaches zero when the bias depletes the channel including the $n=0$ state. Due to the spin splitting by the adjacent magnet, the antiparallel spin state becomes depleted ahead of the parallel state in the conduction band,  while the opposite is the case for the valence band. The consequence thus opens two windows for large polarization with opposite signs. This means that the polarization can be chosen to be either parallel or antiparallel to input magnetization $\mathrm{{\bf M}_1}$ by a simple switch of the applied potential, achieving the desired COPY and NOT operations between two neighboring cells with electrical control as shown in Fig.~\ref{fig_condr}(b).

In the numerical calculation, the magnets are assumed to possess identical properties with the size of $60\times 60 \times 2$~{nm}$^3$, saturation magnetization $\vert \mathbf{M}_{i}\vert = 160$~Oe, intrinsic magnetic anisotropy of 40 fJ/$\mu$m$^3$ along the in-plane hard axis (e.g., $y$), and a damping factor of 0.1.  The exchange coupling energy is taken to be 40 meV at the interfaces (both TI/FM and FM/grahene),~\cite{Semenov2012}  inducing spin splitting of 80 meV for the spin dependent barrier.  The intrinsic chemical potential in graphene and the back gate capacitance are set at 0.3 eV and 0.05 F/m$^2$, respectively.  The signal current on the graphene channel also assumes the strength of 0.1 $\mu$A/nm. These parameters are used hereafter unless explicitly specified otherwise.

\begin{figure}
\includegraphics[width=8.0cm]{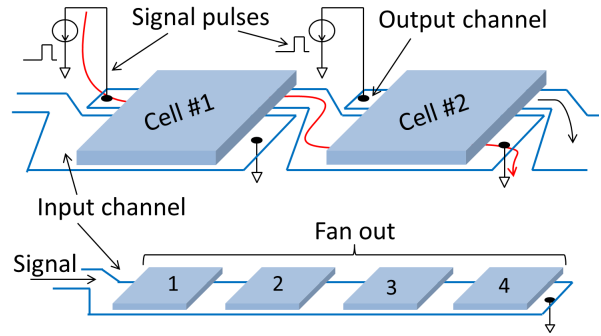}
\caption{(Color online)
Separate input and output channels are defined for the information flow. The red curved arrow indicates the electron path between two neighboring cells.  The fan-out is realized by placing multiple target cells along the same input channel.
}
\label{fig_iofun}
\end{figure}

One potential concern for this scheme based on the spin polarized current is the backward propagation of information, i.e., how to ensure that the spin polarized electrons flow only downstream. This can be addressed by patterning the graphene interconnect to separate the input and output channels as indicated in Fig.~\ref{fig_iofun}. The output of the upstream cell is connected to the input in the downstream, forming an electron path indicated by the red curved arrow, where only the cells in the two neighboring stages are connected at a time. The asymmetric pattern with a wider input channel is to maximize the area of exchange interaction with the target magnet.
For the fan-out, multiple cells can be placed along the same input channel to share the input, whose capacity is limited by electron spin relaxation. Given a typical relaxation length of $4~\mathrm{\mu m}$ in graphene,~\cite{Han2012} the upper limit may approach approximately $40$.
According to the reliability analysis discussed later in the paper (Section~\ref{sec5}), the energy consumption for each COPY/NOT operation could be of the order of \textit{femtojoules} that is dominated by the Joule heating from the signal current.

\subsection{Via electron mediated exchange interactions}

\begin{figure}
\includegraphics[width=7.0cm]{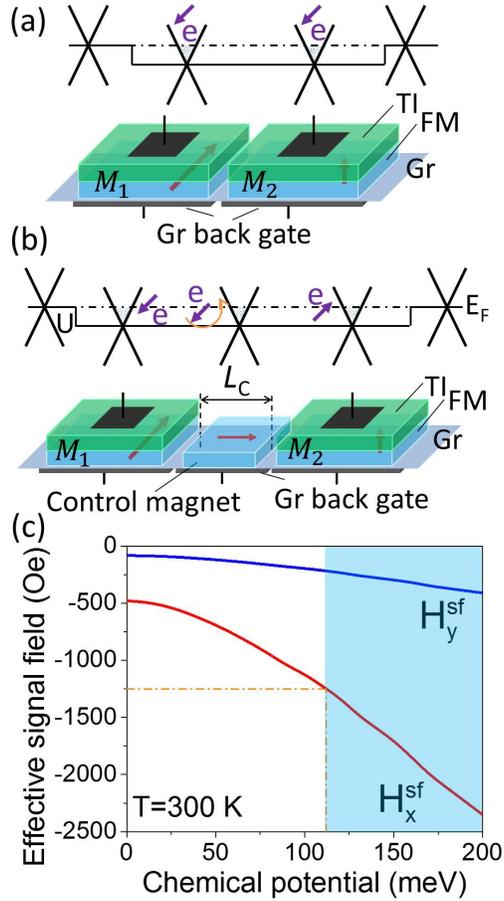}
\caption{(Color online)
(a) COPY  and (b) NOT operations in the current-less approach that are achieved via electrostatically controlled coupling between magnets when the downstream cell is in the Active state.  An energy well can be introduced by simultaneously applying a bias to the graphene back gates. The gap between the gates can be ignored so long as it is smaller than the screening length which is typically several tens of nanometers in graphene.
(c) Effective signal field exerted on the downstream cell as a function of chemical potential in the graphene channel. The shaded region indicates the condition for error rate below $10^{-6}$.
}
\label{fig_smc}
\end{figure}

In contrast to the above approach, a fully electrostatic mechanism can eliminate the Joule heating and thus reduce the power requirement.  As mentioned earlier, the magnetic susceptibility
of graphene electrons can be modulated by a gate bias $-$ a consequence of the linear dispersion relation.~\cite{Duan2012}  This brings an opportunity to electrostatically turn on/off the effective exchange coupling between adjacent magnets that is mediated by the graphene electrons in the channel.
As shown in Fig.~\ref{fig_smc}, an electron potential well can be generated by the graphene back gate to facilitate the overlap of electron wave functions between the two involved cells.  Qualitatively speaking, the upstream cell with a stable magnetization state ($\mathrm{{\bf M}_1} \| \pm\hat{\mathbf{x}}$) would induce electron spin polarization in the graphene channel that diffuses to the downstream cell.  With negligible decay over the device dimension,\cite{Han2012} this aligns the target cell to realize the COPY operation [Fig.~\ref{fig_smc}(a)]. On the other hand, insertion of a control magnet ($\mathbf{M}_{\mathrm{C}} \| \pm\hat{\mathbf{y}}$) in the middle (magnetized normal to $\mathrm{{\bf M}_1}$) would cause spin precession as indicated in Fig.~\ref{fig_smc}(b). If its length is such that the spin experiences a $180^{\circ}$ rotation when reaching the target cell, the anti-parallel alignment can be achieved for the NOT operation.
The required distance for a $\pi$ turn can be estimated as $L_{\mathrm C}=\pi\hbar v_{F}/2G_0\approx 26~\mathrm{nm}$ following the analysis described in an earlier study (with $G_0=40$~meV as specified above).~\cite{Semenov2007} Note that the obtained $L_{\mathrm C}$ is well within the electron spin mean free path in graphene.

The strength of the electron mediated coupling effect can be obtained by considering the induced change in the free energy of the system. For instance, the thermodynamic potential of graphene electrons is a function of magnetic states as:
\begin{eqnarray}
E_c(\mathbf{m}_1,\mathbf{m}_2)&=&-k_{B}T\sum_{b,k}\ln \left\{ 1+2\frac{\exp
\left( \frac{E_{F}-E_{b,k}}{k_{B}T}\right) }{\left[ 1+\exp \left( \frac{%
E_{F}-E_{b,k}}{k_{B}T}\right) \right] ^{2}}\right.  \notag \\
&&\times \left. \left[ \cosh \frac{\Delta _{k}(\mathbf{m}_{1},\mathbf{m}_{2})}{k_{B}T}-1%
\right] \right\} ,  \label{eq_Emex}
\end{eqnarray}%
where $E_F$ is the chemical potential, $b$ ranges the conduction and valence bands and $2 \Delta _{k}(\mathbf{m}_{1},\mathbf{m}_{2})$ corresponds to spin splitting in the graphene band caused by the exchange interaction with the magnets. For convenience, the normalized magnetization vector (i.e., $\mathbf{m}_{i} = \mathbf{M}_{i}/\vert \mathbf{M}_{i}\vert$) is used as all magnets are assumed to have the same saturation magnetization. The expression clearly shows
that the finite spin splitting always decreases the thermodynamic potential.  Thus, the problem of finding the minimum $E_{c}(\mathbf{m}_1,\mathbf{m}_2)$ reduces to a search for the maximum energy splitting $\Delta_{k}(\mathbf{m}_1,\mathbf{m}_2)$.   Subsequent calculations illustrate that the $\mathbf{m}_{2} = \mathbf{m}_{1}$ state indeed provides the minimum $E_{c}$ in Fig.~\ref{fig_smc}(a) (COPY), while it is a state near $\mathbf{m}_{2} = - \mathbf{m}_{1}$ in Fig.~\ref{fig_smc}(b) (NOT).~\cite{SUPP}

The dependence of the free energy on the magnetization of the target cell ($\mathbf{M}_{2}$) may be best interpreted in terms of an effective field that determine its stable state.  Adopting an approach commonly used in the magnetic system, the macroscopic field may be obtained as $ \mu_0 \mathbf{H}^{\rm sf} = - \frac{1}{V_2}\partial E_{c} / \partial {\mathbf{M}_2}$ (where $V_2$ is the volume of the target magnet and $\mu_0$ the permeability constant) that formally defines the orientation and strength of the spin signal.  Figure~\ref{fig_smc}(c) shows the calculated outcome for the NOT gate configuration ($L_{\mathrm C} =26$~nm) as a function of $E_F$.
The presence of both  $x$ and $y$ components indicates that the energy minimum occurs slightly away from the antiparallel $\mathbf{m}_{2} = - \mathbf{m}_{1}$ state.  Nonetheless, the carrier mediated exchange interaction achieves inversion of the spin signal (i.e., polarization) as desired.
The observed enhancement of the signal strength with $E_F$ ensures a robust performance against thermal noise with an applied back gate bias.  For instance, ${H^{\rm sf}}$ of approx.\ 1250 Oe can realize the error rate below $10^{-6}$ (see Sec.~\ref{sec5} for a more detailed discussion).  The necessary shift of 0.1 eV in $E_F$ from the Dirac point translates to the back gate voltage swing of about 0.12 V. The corresponding energy requirement is approximately 10 aJ per COPY/NOT operation including the amount needed to prepare the target cell in the Bennet clocking scheme.  Moreover, with only capacitor charging/discharging, a significant portion of this energy can be recovered in the clock network.\cite{Chan2006}  Accordingly, the net consumption may be reduced to the attojoule level.  In this scheme, it is advantageous to have an intrinsically depleted graphene channel unlike the mechanism based on the spin polarized current.

\section{Construction of logic circuits} \label{sec4}
Once the elemental cell and the cell-to-cell COPY/NOT operations are established, the rest of the Boolean logic can be built on the spin logic platform with majority gates.~\cite{Behin-Aein2010,Kiermaier2013,Imre2006,Nikonov2012}
We demonstrate the logic realization with a 1-bit full adder. It is important to note that the 1-bit adder logic can be decomposed to two majority logic components: the carry-out bit ($\mathrm{C_{out}}$) is the majority gate of inputs $a$,$b$,$c$, and the sum bit (S) equals to a five-input majority logic of $a$,$b$,$c$, and two $\overline{c_\mathrm{out}}$~'s.~\cite{SUPP}

\begin{figure*}
\includegraphics[width=16cm]{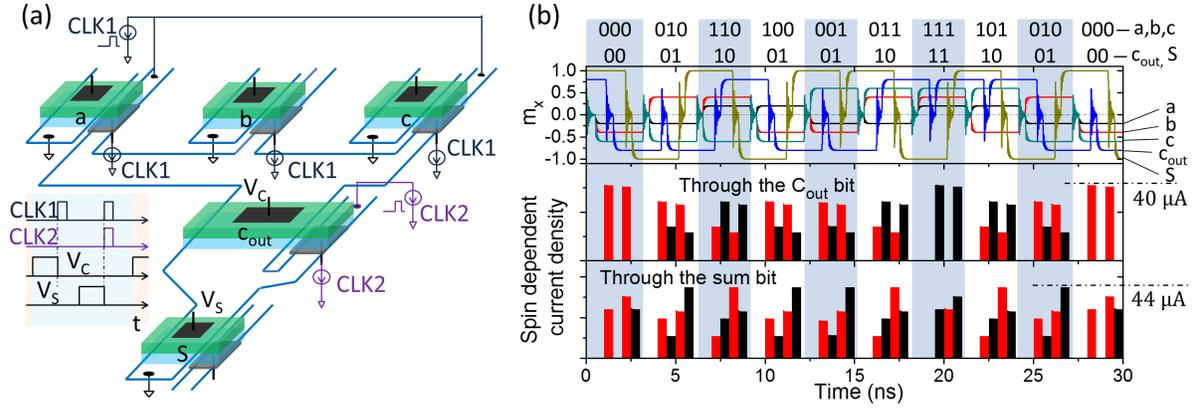}
\caption{(Color online)
(a) Circuit layout of a 1-bit full adder. The inserted shows the clocks and control signals. The graphene back gate biases and the channel inputs share the same clock but may differ in values.
A doubled strength of the outgoing signal from C$_\mathrm{out}$ (specifically, $\overline{c_\mathrm{out}}$) can be achieved by adjusting the cell size or the channel driving voltage.
(b) Results of ten add operations performed with the input states set dynamically in the simulation.  The top panel shows the magnetization of each cell and the bottom two provide the spin parallel ($\hat{\mathbf{x}}$, black/darker) and spin anti-parallel ($- \hat{\mathbf{x}}$, red/lighter) currents through the input channels of C$_\mathrm{out}$ and S, respectively. The magnetization $m_x$ varies between $1$ and $-1$ in all five cases (e.g., $m_x= \pm 1$ for logic "1" and "0", respectively).  The heights are adjusted artificially to distinguish the curves from each other for easier viewing (top panel).  Similarly, the spin anti-parallel current is artificially shifted to the left by 0.5 ns to separate it from the spin parallel current (bottom panels).
}
\label{fig_adder}
\end{figure*}

Following the spin current based interconnect scheme (Sec.~III.A), Fig.~\ref{fig_adder}(a) shows a 1-bit adder layout to accomplish the two-stage operations.  In the first stage, electrons are injected by clock CLK1 through $a$,$b$,$c$ cells to set the state of $\mathrm{C_{out}}$. The second stage includes another clock CLK2 to inject electrons through the output channel of $\mathrm{C_{out}}$ (the narrow channel) with the reversed polarization (i.e., $\overline{c_\mathrm{out}}$). Doubling the $\overline{c_\mathrm{out}}$ signal can be achieved by either adjusting the cell size or increase the driving voltage in the corresponding output channel. At the circuit level, each unit acts as a set of resistors and the circuit energy consumption scales with the logic complexity. We develop a fully coupled device-circuit simulation method~\cite{SUPP} to verify the adder behavior with ten successive add operations that covers the truth table [Fig.~\ref{fig_adder}(b)].  In this simulation, we assume that all the magnets have the same parameters as previously stated and the input/output channels are divided with a ratio of 4:1. The driving voltage is set to 0.3 V except the output channel of $\mathrm{C_{out}}$ that is 0.45 V to double the signal strength. The back gate voltages are chosen according to Fig.~\ref{fig_condr}; i.e., $-0.45$ V for COPY and $-0.6$ V for NOT.  As each stage of magnetization switching is achieved within the period of 1 ns,  a total of 3 ns is needed for the 1-bit adder including including the process to prepare the input states $a$,$b$,$c$.  With the specified conditions, the total channel current is around $30\sim 80$~$\mu$A and the corresponding energy consumption approx.\ 15~fJ.

\begin{figure}
\includegraphics[width=8.0cm]{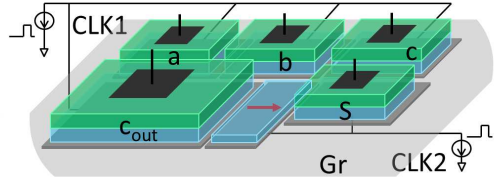}
\caption{(Color online)
Tentative circuit layout for the 1-bit adder in the current-less approach. It uses the same control clocks and follows the 2-stage operation as in the current driven design.
}
\label{fig_addert}
\end{figure}

When the current-less operating mechanism is adopted (Sec.~III.B), on the other hand, construction of the logic circuits essentially amounts to arranging each elemental cell on a universal graphene sheet with properly clocked gates. It follows a different methodology compared to the current based circuits. One characteristic is the compact layout required by the nature of local exchange interactions. A tentative 1-bit adder design is shown in Fig.~\ref{fig_addert} following the same two-stage operating procedure. In the first stage, CLK1 induces the electron wave function overlap in the graphene channel between the input cells and output cell $\mathrm{C_{out}}$ to achieve a 3-input majority gate. The next stage has CLK1 and CLK2 applied together to overlap all the cells for the equivalent 5-input majority logic. The inserted control magnet inverts the signal from ${c_\mathrm{out}}$ to $\overline{c_\mathrm{out}}$.  In the actual implementation, however, this trial design may need adjustments as it is based on the assumption that the outcome from our two-cell analytic prediction holds for the complex geometry at least qualitatively.

\section{Magnetic switching performance and reliability} \label{sec5}

For comprehensive evaluation of the proposed logic devices, it is important to characterize the switching dynamics and verify the robustness in a thermal bath.
Particularly, the Bennett clocking scheme relying on two successive 90$^\circ$ magnetization rotations via a meta-stable state is inherently susceptible to the fluctuations that could limit not only the operation accuracy but also the switching speed.  The well established Landau-Lifschitz-Gilbert (LLG) equation is used to numerically examine these issues.~\cite{SUPP}  The investigation primarily considers the spin current based interconnect scheme with 100\% polarization for simplicity.  The performance of the electrostatic approach can also be understood by correlating the strength of the spin current to the induced effective magnetic field.  In fact, this analysis may be applicable more broadly to other spin logic realizations that utilize similar operating principles.

\subsection{Switching Speed}

As described in Sec.~\ref{sec2}, Fig.~\ref{fig_cell}(c) illustrates a snapshot of switching with Bennet clocking. One particularly interesting parameter that warrants additional scrutiny is the hard-axis anisotropy $K_y$ along the $y$ axis as it can significantly influence the details of rotational dynamics.  The expectation is that the magnetization could switch and relax faster with a larger $K_y$ since it tends to confine the switching path to the $x$-$z$ plane. Our simulation of the first 90$^{\circ}$ rotation (with $m_{x}$ changing from $1$ to $0$ via the top gate bias) indeed indicates that the desired operation can be achieved more quickly with the characteristic time shorter than $0.5$~ns once $K_y$ increases above approx.\ 20$-$30 fJ/$\mu$m$^3$.
It should be noted that  application of the hard-axis anisotropy in the $y$ direction together with the demagnetization field amounts essentially to "the easy axis" along the ${x}$ direction, which has been assumed by numerous studies in the literature.~\cite{Behin-Aein2010,Augustine2011}  Roughly speaking, the hard-axis anisotropy must be at least comparable to the demagnetization energy for the desired confinement effects on the switching dynamics.

\begin{figure}
\includegraphics[width=7.0cm]{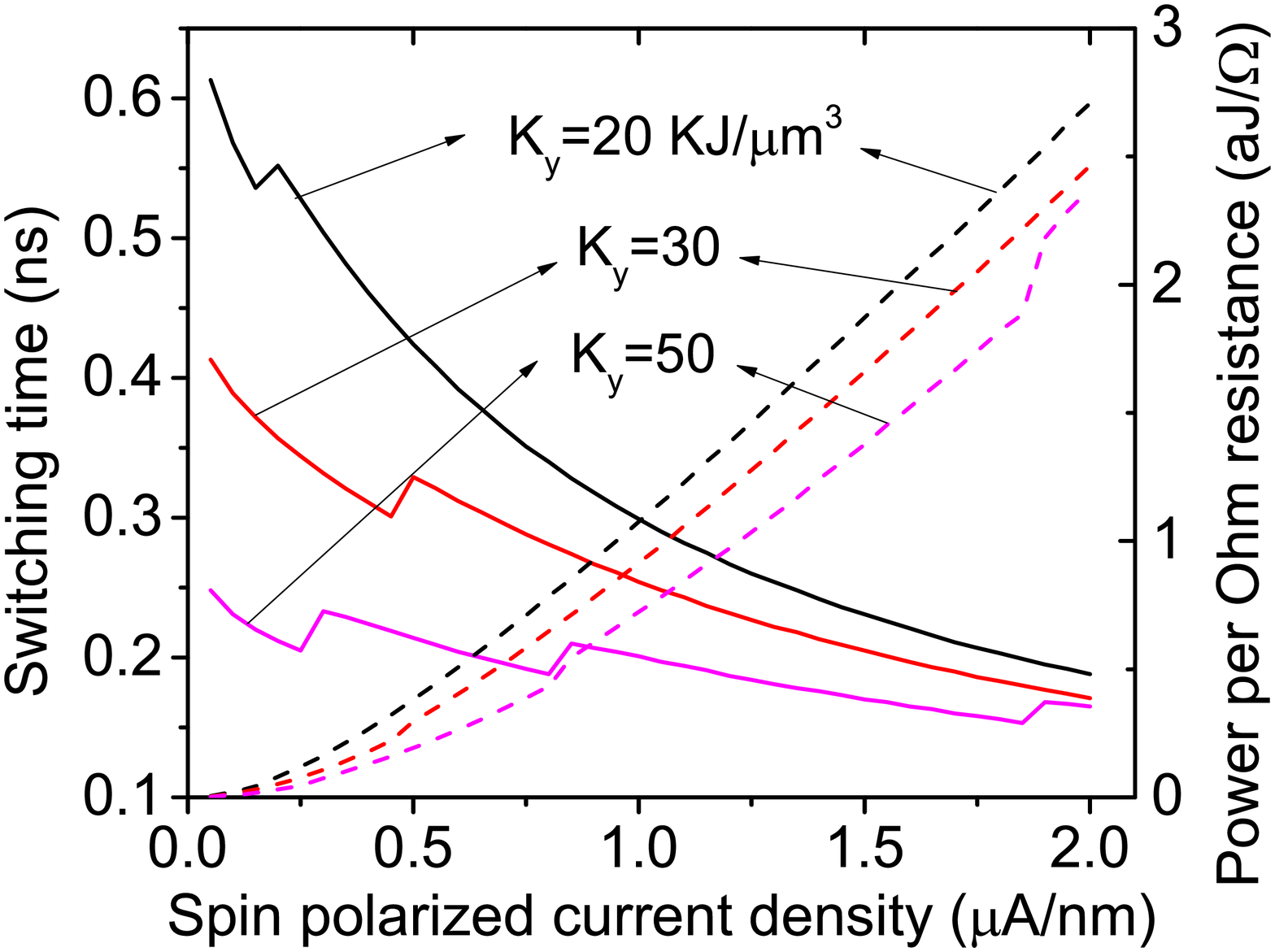}
\caption{(Color online)
Switching time (solid lines) and corresponding power consumption (dashed lines) of the current driven relaxation process from $m_{z}=1$ to $m_{x}=1$ ($z \rightarrow x$). The relaxation is marked completed when $|m_{x}|> 0.9$. The power consumption per ohm is calculated as $I^{2}t$, where $I$ is the total current assuming a $60$-nm channel width and $t$ is the duration of relaxation. }
\label{fig_tktj}
\end{figure}

The second half of the full $180^{\circ}$ rotation is also improved by a larger hard-axis anisotropy as indicated in Fig.~\ref{fig_tktj}. Here, a continuous signal current instead of a pulse [i.e., Fig.~\ref{fig_cell}(c)] is considered to drive the magnetization until it reaches the correct direction to capture the main features. Both the polarized current and the hard-axis anisotropy provide the driving force for relaxation, and the switching time drops when their values increase. At a small current density, the relaxation is mainly driven by the intrinsic anisotropy field so that the switching time varies significantly over different anisotropy values. As the current rises, the exchange torque also drives the switching process to reduce the switching time. When the current is high enough to dominate over the contribution from the anisotropy, the curves tend to converge (to approximately 0.1$-$0.2 ns). It is also worth noting that at a sufficiently large anisotropy (e.g., $K_{y} \gtrsim 50$ fJ/$\mu$m$^3$), the influence of signal amplitude  becomes insignificant in the simulation range. The corresponding dashed lines indicate the energy consumption per ohm resistance. It shows a steep increase for switches with a high current even though the duration reduces.  Accordingly, a large $K_y$ appears to be generally favorable (i.e., for both fast switching and low energy consumption).


\subsection{Possible error sources}
\begin{figure*}
\includegraphics[width=8.0cm]{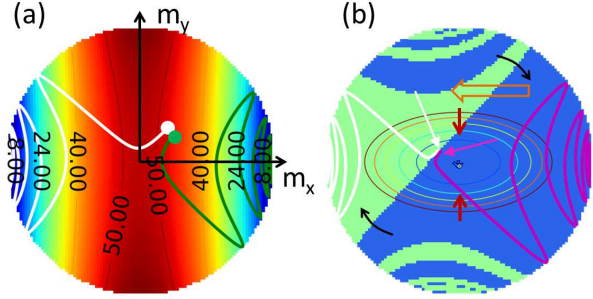}
\caption{
Switching characteristics related to the operation error rates. (a) Intrinsic energy landscape of the magnet that drives the relaxation after the signal pulse.  The green line shows a path along which the magnetization relaxes to the same side of the initial state, while the white one indicates that a low energy valley can lead the magnetization to the opposite side. (b) Topology of switching success/failure with a signal current pulse of 0.2~ns at 0.2~$\mu$A/nm. The null-state magnetization that relaxes to $m_{x}=1$ is shown in blue (success), while that led to $m_{x}= -1$ is marked in green (failure).  The white and purple curves provide two sample paths (with the arrows pointing the starting locations). The dot in the middle indicates the magnetization distribution at the end of a $0.5$-ns biasing stage. The ellipses indicate the lowest energy contours with an increment of 2${k_{B}T}$ from the meta-stable state.}
\label{fig_path1}
\end{figure*}

In the magnetic switching process based on Bennett clocking, the switching errors are mainly caused by deviation from the meta-stable state (after the first $90^{\circ}$ rotation) as well as the low energy path leading to the energy minimum with unintended polarization. One such example is illustrated in Fig.~\ref{fig_path1} based on the magnetization phase-space analysis.  As displayed by the white and green curves in Fig.~\ref{fig_path1}(a), two states with nearly identical locations in the intrinsic free energy landscape (close to the meta-stable $m_z= 1$) can end up with two drastically different relaxation pathes to opposite polarizations.  This can be attributed to the convoluted and close entanglement in the high energy regions in
the context of precessional dynamics.~\cite{Bertotti2003} Evidently, the rotational nature of magnetization switch adds complexities to the problem.  Increasing the signal intensity, while certainly helpful, is only part of the solution for the robust performance.  Figure~\ref{fig_path1}(b) summarizes the success/failure of the switching operation when a current pulse of $0.2~\mathrm{ns}$ at $0.2~\mathrm{\mu A/nm}$ is applied to set the final state to $m_{x}=1$.  The darker colored (blue) region represents the initial "null" states from which the magnetization relaxes to the desired final state  with the aid of signal current (i.e., $m_{x}=1$; success), whereas the lighter region (green) results in the failure or error with the final polarization in the opposite direction ($m_{x}= -1$).  An increased signal intensity moves the lighter/darker boundary  towards the failure side (i.e., less failure; see the block arrow).  The larger hard-axis anisotropy $K_y$, on the other hand, rotates the boundary clockwise (see the black arrows).  Aside from the asymmetric pattern of the switching map, the concentric ellipses show the constant energy contours with an increment of 2${k_{B}T}$ from the meta-stable state.  This gives a qualitative measure of random thermal fluctuation.  For instance,  the  trouble spot in Fig.~\ref{fig_path1}(b) would be the lighter colored region within a given ellipse (leading to an error).  In the present discussion, only the states with $m_z \geq 0$ are considered with its value determined by $\sqrt{1- m_x^2 - m_y^2}$ in the 2D plot; the pattern for $m_z\leq0$ satisfies the reflection symmetry.

To be more precise, two major sources can lead to the unintended spread in the null state distribution after the initial 90$^\circ$ rotation; namely, insufficient relaxation and thermal fluctuations. The former would dominate only if the operating frequency is too high, while the later always exists at the level of severity determined by the temperature. According to Sec.~\ref{sec5}.A, the switching time to the meta-stable state (i.e., the first $90^\circ$ rotation) is well within 1 ns (e.g., $\lesssim 0.5$~ns), indicating that the insufficient relaxation can easily be avoided.  The case of Fig.~\ref{fig_path1}(b) clearly illustrates this point, where the applied bias of 0.5 ns sufficiently concentrates the distribution to the desired $m_z=1$ state (see the tight distribution near the center). Consequently, the distribution of null state magnetization in a well designed operating condition is determined by thermal fluctuations and errors would occur if the signal is not able to remedy all of the possible magnetization within this distribution including the added complexities in the precessional dynamics.

\subsection{Error rate evaluation}
\begin{figure}
\includegraphics[width=8.0cm]{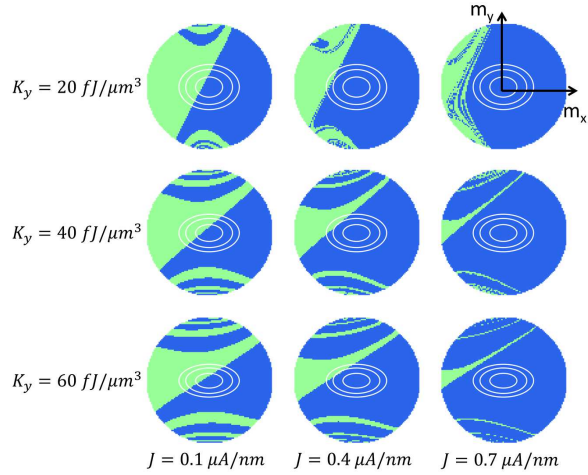}
\caption{
Topology of switching success/failure with a signal current pulse of 0.4~ns in duration at different values of strength $J$ and anisotropy $K_y$. The null-state magnetization that relaxes to $m_{x}=1$ is shown in blue (success), while that leads to $m_{x}= -1$ is marked in green (failure). The magnet has a dimension of $60\times 60\times 2$~nm$^3$. The ellipses indicate the energy contours of 2$k_{B}T$, 4$k_{B}T$, and 10$k_{B}T$ from the meta-stable state.}
\label{fig_succmaps}
\end{figure}

The device robustness is closely related to the switching details.  A conventional treatment to examine the  performance in a realistic environment is to add a white thermal field to the LLG equation (i.e., the stochastic LLG equation) that induces a Brownian motion by virtue of the correlation assumptions;~\cite{Brown1963,Cimrak2008} i.e., $\mathbf{H}_\mathrm{eff}^{'}= \mathbf{H}_\mathrm{eff}+\mathbf{H}_\mathrm{th}$.
The random thermal field $\mathbf{H}_\mathrm{th}$ is described by a Gaussian distribution with the variance determined from the fluctuation-dissipation theorem:
\begin{equation}
\langle H_{th}^i \left( t_{1}\right)H_{th}^j \left( t_{2}\right)\rangle = \frac{2k_{B}T\alpha}{\mu_0 VM_{0}\gamma} \delta_{i,j} \delta\left( t_{1}-t_{2}\right) ,
\label{eq_Hth}
\end{equation}
where $V$ is the magnet volume and indices $i,j$ correspond to the coordinate axes.~\cite{Brown1963,Berkov2002}  As mentioned, this term can be readily included in the calculation.  One major difficulty of the stochastic approach, however, is that the number of the required simulations increases at least linearly with the desired accuracy.  For instance, the simulations must be repeated $10^{6}$ times or more in order to accurately estimate the error rate of $10^{-6}$ (i.e., one incorrect event out of $10^6$ operations), not to mention the numerical complexities associated with various discretization issues in the actual implementation.~\cite{Berkov2007}  For an alternative, computationally more efficient method, it is worth noting that the magnet is most vulnerable to thermal fluctuation at the null state.  Accordingly, we consider the thermal variation/noise explicitly through the null state magnetization distribution, while the relaxation dynamics is treated deterministically based on the LLG equation.  Then, the error rate $P_s$ can be estimated as:
\begin{equation}
P_{s}= 1 - \int_{m_z \geq 0} d{\mathbf{m}_\mathrm{n}} {R}(\mathbf{m}_\mathrm{n}) F(\mathbf{m}_\mathrm{n}) / \int_{m_z \geq 0} d{\mathbf{m}_\mathrm{n}} F(\mathbf{m}_\mathrm{n})  ,
\label{eq_Psucc}
\end{equation}
where $F(\mathbf{m}_\mathrm{n})$ is the distribution of the null state magnetization $\mathbf{m}_\mathrm{n}$ and ${R}(\mathbf{m}_\mathrm{n})$ denotes the simulated switching result. More precisely, ${R}(\mathbf{m}_\mathrm{n})=1$ if the operation results in the desired outcome and ${R}(\mathbf{m}_\mathrm{n})=0$ for the error/failure. Figure~\ref{fig_succmaps} shows the results of ${R}(\mathbf{m}_\mathrm{n})$ on the $x$-$ y$ plane for a number of cases.  The plots clearly illustrate the earlier statement that the boundary between the failure and success regions moves towards the failure side when the signal current $J$ increases and rotates clockwise when the hard-axis anisotropy $K_y$ increases.  As for $F(\mathbf{m}_\mathrm{n})$, we use a Boltzmann distribution to weigh each possible null state magnetization, i.e., $F(\mathbf{m}_\mathrm{n}) =\exp{[-E_{m}(\mathbf{m}_\mathrm{n})/k_{B}T]}$, where $E_{m}$ represents the magnetic energy including the bias induced out-of-plane anisotropy.  This choice can be justified since the thermal fluctuations would be dominant over the variation by insufficient initial relaxation under proper operating conditions (see the discussion in Sec.~\ref{sec5}.B).  Of the iso-energy contours plotted in Fig.~\ref{fig_succmaps}, it is interesting to note that nearly 90\% of the thermal distribution is contained in the first 2$k_BT$.

\begin{figure*}
\includegraphics[width=13.0cm]{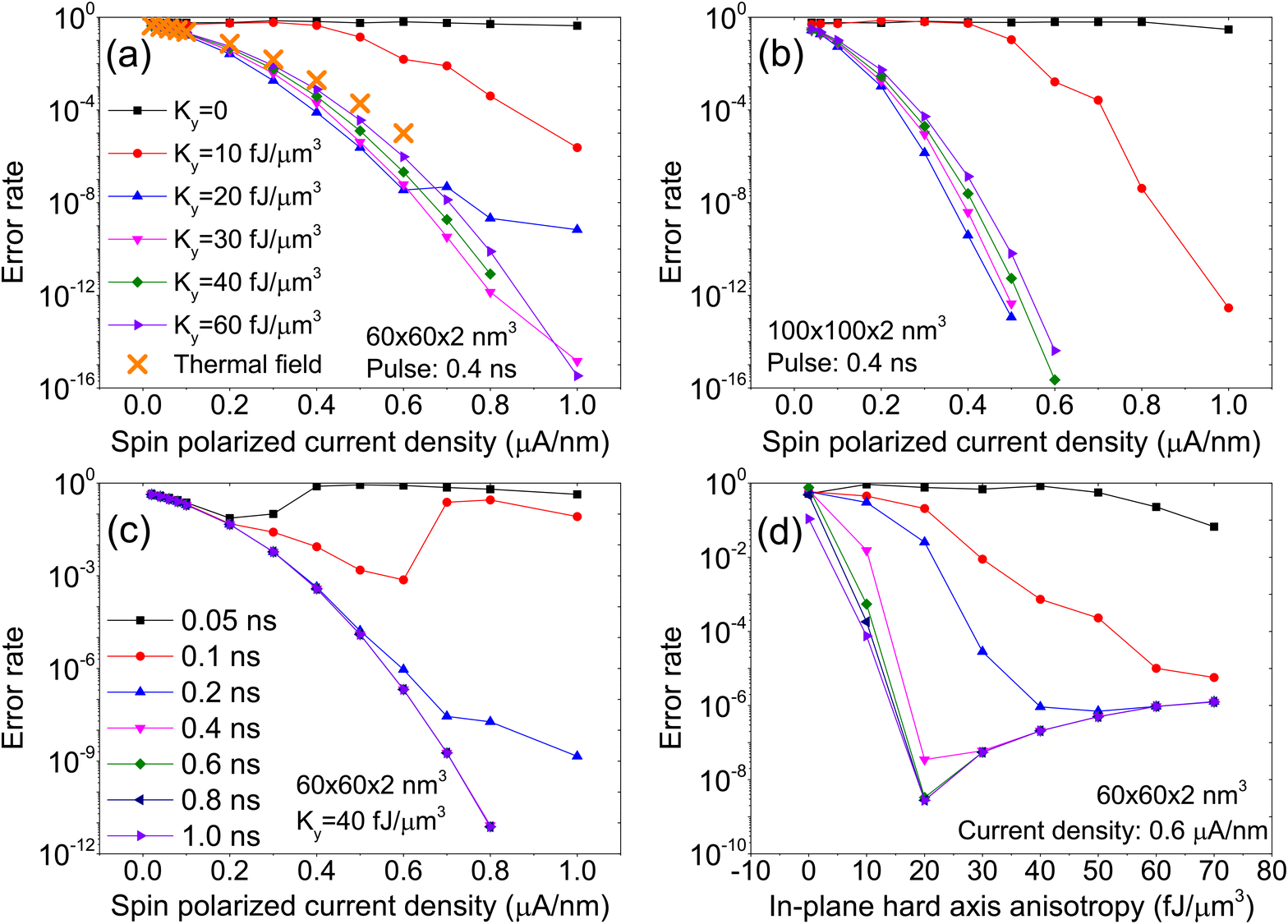}
\caption{
Variation of switching error rate over magnet size, signal current duration, signal current density, and hard-axis anisotropy.  Identical line colors and symbols are used in (a) and (b) to denote different values of $K_y$, whereas (c) and (d) share the same notation on the pulse duration.  In (c), the curves for duration over 0.4 ns essentially overlap with each other.  The orange-colored data points in (a) represent the results from the random field approach with $K_{y}=60$~fJ/$\mu$m$^3$.  All other curves are obtained by considering the thermal distribution of the null-state magnetization.}
\label{fig_ERall}
\end{figure*}

The calculated error rates are shown in Fig.~\ref{fig_ERall} as a function of in-plane hard-axis anisotropy, magnet size, signal pulse duration, and the strength.  From the results, it is evident that the device can reach the desired high degree of robustness once the hard-axis anisotropy becomes sufficiently large. In fact, $K_y$ above approx.\ 20$-$30 fJ/$\mu$cm$^3$ appears to converge without a large deviation between the different values.  Accordingly, the threshold current density for a target error rate (say, $ 10^{-4} \sim 10^{-6}$) is expected to be relatively insensitive to this crucial parameter [see Fig.~\ref{fig_ERall}(a,b)].  When the magnet size increases, the switching becomes generally more reliable; this can be attributed to the larger surface area enabling a stronger interaction with the signal current.  In the case of signal pulse duration, it shows a dependence akin to that of $K_y$. As can be seen from Fig.~\ref{fig_ERall}(c), the proposed device provides very comparable performances once the signal pulse is approx.\ 0.2 ns or longer.  One key difference is that the error rate may exhibit a threshold behavior on the duration.  Namely, there may be a minimum pulse length below which the operation cannot attain high fidelity even with an increase in the signal current strength (see, for example, the case of 0.1 ns).  The requirement on the pulse duration may be partly compensated by a larger in-plane hard-axis anisotropy.  Figure~\ref{fig_ERall}(d) illustrates the point clearly, where the performance of the 0.1-ns case converges to an error rate similar to those of the longer pulses as the anisotropy energy increases beyond the demagnetization terms.  An additional finding of interest in Fig.~\ref{fig_ERall}(d) is that the error rates for the longer pulses ($\gtrsim 0.4$~ns) seem to reach the minimum at around $K_y=20~\mathrm{fJ/\mu m^3}$ and then rise afterward with a converging trend in the end. This can be understood by examining evolution of the boundary discussed in Fig.~\ref{fig_path1}. The clockwise rotation may expose more thermally distributed region to the failure part (note the lack of circular symmetry in the contours) that, combined with a longer pulse, could make the relaxation dynamics less stable. Finally, a comparison is made with the results obtained by the conventional random field.  As illustrated in Fig.~\ref{fig_ERall}(a), both approaches show good agreement for the case of $K_y =60$ fJ/$\mu$cm$^3$.  The accuracy beyond $10^{-5}$ cannot be addressed in the white field treatment due to the limited number of simulation repeats ($10^5$).

The reliability analysis given above verifies the feasibility of the proposed devices. We can thus reasonably set the switching period to 1 ns, with a 0.5-ns TI gate bias followed by a 0.5-ns signal pulse, which is the chosen condition for the 1-bit adder simulation (see Sec.~\ref{sec4}).  The results also indicate a room for further improvement with a total $180^{\circ}$ switching period as short as 0.5 ns.  The sub-nanosecond switching time is crucial in achieving low energy consumption, while maintaining sufficient fidelity of operation.  In addition, the built-in non-volatility appears attainable with little or no overhead to the performance specifications.  The estimated free-energy barrier of the magnet under discussion ($60\times 60\times 2$~nm$^3$, 30~fJ/$\mu$m$^3$) is well over 40${k_{B}T}$.

\section{Conclusion}
The theoretical analysis based on the LLG equation clearly demonstrate that the nanomagnet-Dirac fermion heterostructures provide a unique environment to realize the long desired goal of inducing and controlling, by electrical means, strongly correlated interactions between electronic and magnetic systems.  The proposed device concepts offer a promising alternative in the development of post-CMOS, low-power devices.  Further, it is worth noting that application of the spin logic can go beyond the conventional Boolean architecture.  With the coupling dependent on the local ensemble of electrons, structures resembling cellular automata and/or neural network may potentially be achieved as well.

%

\section*{Acknowledgements}
This work was supported, in part, by the US Army Research Office and FAME (one of six centers of STARnet, a SRC program sponsored by MARCO and DARPA).

%

\clearpage

\clearpage

\clearpage

\clearpage

\clearpage

\clearpage

\clearpage

\clearpage

\clearpage


\clearpage

\begin{thebibliography}{}

\bibitem{Chiba2008} D. Chiba, M. Sawicki, Y. Nishitani, Y. Nakatani, F. Matsukura, and H. Ohno, Nature {\bf 455}, 515 (2008).
\bibitem{Wang2013} K. L. Wang,  J. G. Alzate, and P. K. Amiri, J. Phys. D: Appl. Phys. {\bf 46}, 074003 (2013).
\bibitem{Pesin2012} D. Pesin and A. H. MacDonald, Nat. Mater. {\bf 11}, 409 (2012).
\bibitem{Hasan2010} M. Z. Hasan and C. L. Kane,  Rev. Mod. Phys. {\bf 82}, 3045 (2010).
\bibitem{Han2012} W. Han,  K. M. McCreary, K. Pi, W. H. Wang, Y. Li, H. Wen, J. R. Chen, and R. K. Kawakami, J. Magn. Magn. Mater. {\bf 324}, 369 (2012).
\bibitem{Chen2010}  Y. L. Chen, J.-H. Chu, J. G. Analytis, Z. K. Liu, K. Igarashi, H.-H. Kuo, X. L. Qi, S. K. Mo, R. G. Moore, D. H. Lu, M. Hashimoto, T. Sasagawa, S. C. Zhang, I. R. Fisher, Z. Hussain, and Z. X. Shen,  Science {\bf 329}, 659 (2010).
\bibitem{Kou2012} X. F. Kou, W. J. Jiang, M. R. Lang, F. X. Xiu, L. He, Y. Wang, Y. Wang, X. X. Yu,
A.V. Fedorov, P. Zhang, and K. L. Wang,  J. Appl. Phys. {\bf 112}, 063912 (2012).
\bibitem{Schlenk2013} T. Schlenk, M. Bianchi, M. Koleini, A. Eich, O. Pietzsch, T. O. Wehling,
T. Frauenheim, A. Balatsky, J.-L. Mi, B. B. Iversen, J. Wiebe, A. A. Khajetoorians, P. Hofmann,
and R. Wiesendanger,  Phys. Rev. Lett. {\bf 110}, 126804 (2013).
\bibitem{Zhang2012d} J.-M. Zhang,  W. Zhu, Y. Zhang,  D. Xiao, and Y. Yao, Phys. Rev. Lett. {\bf 109}, 266405 (2012).
\bibitem{Kong2011} B. D. Kong,  Y. G. Semenov, C. M. Krowne, and K. W. Kim,  Appl. Phys. Lett. {\bf 98}, 243112 (2011).
\bibitem{Yokoyama2011} T. Yokoyama, Phys. Rev. B {\bf 84}, 113407 (2011).
\bibitem{Haugen2008a} H. Haugen, D. Huertas-Hernando, and A. Brataas, Phys. Rev. B {\bf 77}, 115406 (2008).
\bibitem{Yang2013} H. X. Yang, A. Hallal, D. Terrade, X. Waintal, S. Roche, and M. Chshiev, Phys. Rev. Lett. {\bf 110}, 046603 (2013).
\bibitem{Swartz2012} A. G. Swartz, P. M. Odenthal, Y. Hao, R. S. Ruoff, and R. K. Kawakami, ACS Nano {\bf 6}, 10063 (2012).
\bibitem{Duan2012} X. Duan, V. A. Stephanovich, Y. G. Semenov, and K. W. Kim,  Appl. Phys. Lett. {\bf 101}, 013103 (2012). 
\bibitem{Ji2012} H. Ji, J. M. Allred, N. Ni, J. Tao, M. Neupane, A. Wray, S. Xu, M. Z. Hasan, and R. J. Cava, Phys. Rev. B \textbf{85}, 165313 (2012).



\bibitem{Bennett1982a} C. H. Bennett, Int. J. Theor. Phys. {\bf 21}, 905 (1982).
\bibitem{Semenov2012} Y. G. Semenov, X. Duan, and K. W. Kim,  Phys. Rev. B {\bf 86}, 161406 (2012).
\bibitem{Akhmerov2008} A. R. Akhmerov and C. W. J. Beenakker, Phys. Rev. B {\bf 77}, 085423 (2008).
\bibitem{Semenov2007} Y. G. Semenov, K. W. Kim, and J. M. Zavada,  Appl. Phys. Lett. {\bf 91,} 153105 (2007).
\bibitem{SUPP} See the correspnding sections of Supplementary at [URL will be inserted later] for addional details.
\bibitem{Chan2006} S. C. Chan,  K. L. Shepard, and P. J. Restle, IEEE J. Solid-State Circuits {\bf 41}, 2083 (2006).
\bibitem{Behin-Aein2010} B. Behin-Aein, D. Datta, S. Salahuddin, and S. Datta,  Nat. Nanotechnol. {\bf 5}, 266 (2010).
\bibitem{Imre2006}  A. Imre, G. Csaba, L. Ji, A. Orlov, G.H. Bernstein, and W. Porod,  Science {\bf 311}, 205 (2006).
\bibitem{Kiermaier2013} J. Kiermaier, S. Breitkreutz, I. Eichwald, M. Engelstädter, X. Ju, G. Csaba,
D. Schmitt-Landsiedel, and M. Becherer,  J. Appl. Phys. {\bf 113}, 17B902 (2013).
\bibitem{Nikonov2012} D. E. Nikonov, and I. A. Young,  in {\it Proc. Int. Electron Device Meeting}, 2012, pp.\ 25.4.1-25.4.4.
\bibitem{Augustine2011} C. Augustine, G. Panagopoulos, B. Behin-Aein, S. Srinivasan, A. Sarkar,
and K. Roy, in {\it Proc. IEEE/ACM Int. Symp. Nanoscale Archit. (NANOARCH)}, 2011, pp.\ 129-136.



\bibitem{Bertotti2003} G. Bertotti, I. D. Mayergoyz,  C. Serpico, and M. d'Aquino,  IEEE Trans. Magn. {\bf 39}, 2501 (2003).
\bibitem{Brown1963} W. F. Brown,  Phys. Rev. {\bf 130}, 1677 (1963).
\bibitem{Cimrak2008} I. A Cimr\'{a}k, Arch. Comput. Method. Eng. {\bf 15}, 1 (2008).
\bibitem{Berkov2002} D. V. Berkov, IEEE Trans. Magn. {\bf 38}, 2489 (2002).
\bibitem{Berkov2007} D. V. Berkov,  in {\it Handbook of Magnetism and Advanced Magnetic Materials}, edited by H. Kronm\"{u}ller and S. Parkin (Wiley, New York, 2007), Vol.\ 2, pp.\ 1-29.

\end{thebibliography}
\end{document}


\title{Supplementary: Spin Logic via Controlled Correlation in Nanomagnet-Dirac Fermion Heterostructures}

\author{Xiaopeng Duan}
\affiliation{Department of Electrical and Computer Engineering, North Carolina State University, Raleigh, NC 27695-7911}

\author{Yuriy G. Semenov}
\affiliation{Department of Electrical and Computer Engineering, North Carolina State University, Raleigh, NC 27695-7911}

\author{Ki Wook Kim}
\email{kwk@ncsu.edu}
\affiliation{Department of Electrical and Computer Engineering, North Carolina State University, Raleigh, NC 27695-7911}
\maketitle


\tableofcontents 

\section{Equivalent circuit model of the device} \label{sect_circ}

\begin{figure}
\includegraphics[width=4.5cm]{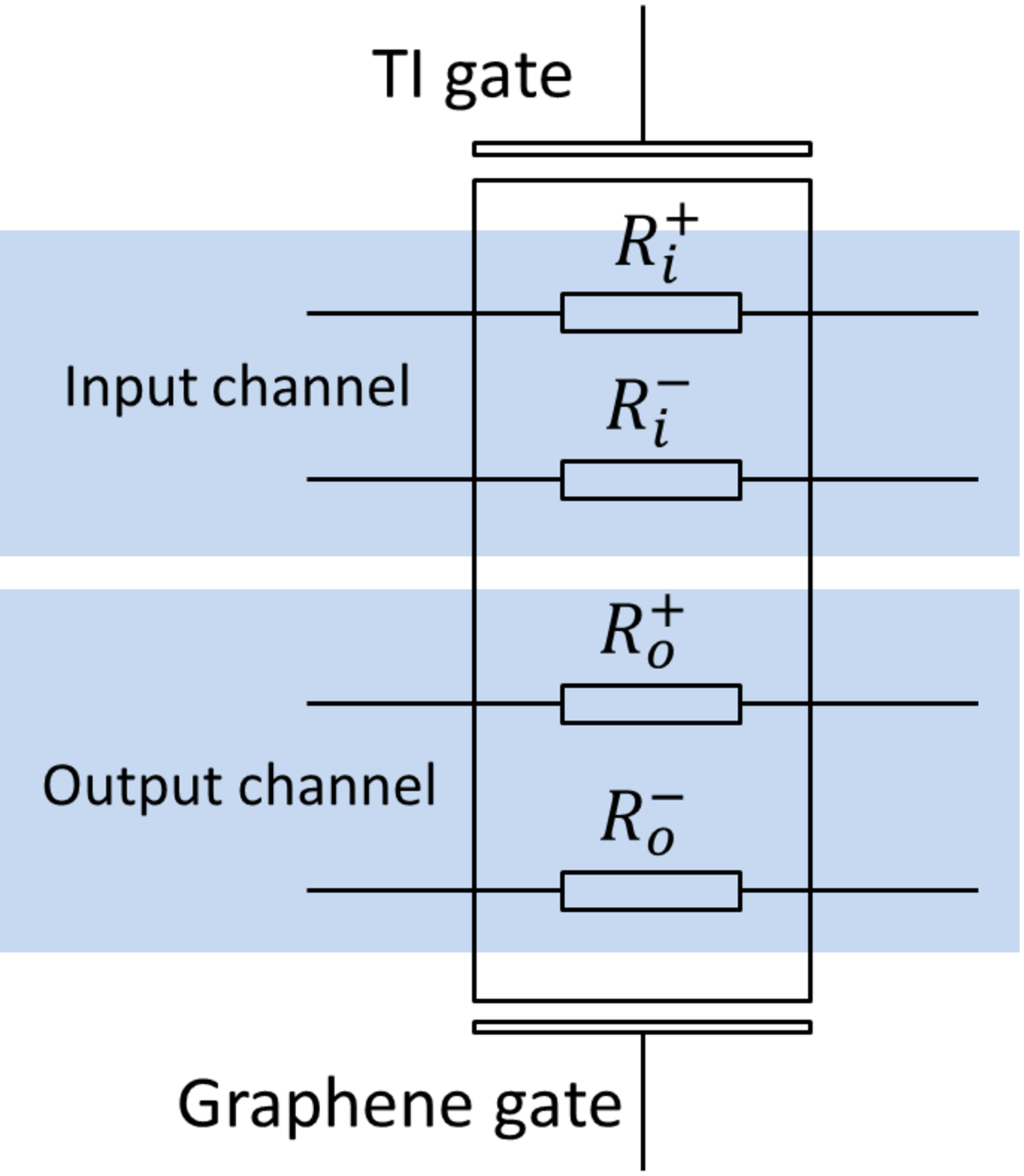}
\caption{
Circuit model for one device cell. Two spin ($\pm 1/2$) channels are considered independent for both input and output, resulting in 4 resistors. The resistances are determined by the bias applied to the graphene back gate and the magnetization state.  The back gate bias is synchronized with the signal current.  The bias pulse applied to the TI top gate in a Bennett clocking scheme controls the $90^{\circ}$ rotation.  The magnetization dynamics is solved together with the transient circuit simulation.}
\label{fig_unitcir}
\end{figure}

A circuit model is developed to assess the circuit performance of the device prototype with a number of simplifying assumptions.
More specifically, each device is treated as a set of bias and magnetization controlled resistors. Spin relaxation in the graphene interconnect is not considered due to the short length/transit time between the two neighboring cells.  Hence, two spin channels (i.e., $+1/2$ and $-1/2$) are considered independent. This leads to the equivalent device model shown in Fig.~\ref{fig_unitcir} that depicts a 10-terminal abstraction consisting of four resistors and two gates. The resistors represent the resistance for spin $+1/2$ ($-1/2$) electrons in the input $R_{i}^{+}$ ($R_{i}^{-}$) and output $R_{o}^{+}$ ($R_{o}^{-}$) channels; see Fig.~3 in the main paper for the interconnect layout. The resistance values are determined by Eq.~(2) in the main paper as a function of the magnetization state and the bias potential barrier. The graphene back gate is applied to directly control the output (reading) channel resistance ($R_{o}^{+,-}$), while the input (writing) channel is unaffected by this electrode as indicated in Fig.~5 (main paper). The TI (i.e., top) gate impacts the resistance through modulation of the magnetization state. The applied bias pulses following the Bennett clocking scheme rotate the magnetization easy axis from in-plane to out-of-plane.

\section{Logic circuit of a 1-bit adder}

Spin/magnetic device circuits are often built based on majority logic gates.~\cite{Behin-Aein2010,Yao2012a,Nikonov2012} For a general majority gate accepting $2l+1$ inputs, the logic output $M_{2l+1}(a_{1}, a_{2}, ...\ a_{2l+1})$ can be expressed as the union of all possible intersections of $l+1$ inputs. Specially for a 3-input majority gate, we have $M_{3}(a_{1},a_{2},a_{3})=a_{1}a_{2}+a_{1}a_{3}+a_{2}a_{3}$. In a 1-bit full adder with inputs $a$,$b$,$c$, the output of the carry-out bit $c_\mathrm{out}$ is $ab+ac+bc$ that is exactly a majority logic as $M_{3}(a,b,c)$. The sum bit $S$ has also been identified~\cite{Augustine2011} as a majority gate logic with 5 inputs consisting of $a,b,c$ and two $\overline{c_\mathrm{out}}$~'s. It can be proven with Boolean algebra as following:
\begin{equation}\label{eq_S}\begin{split}
S&= abc+a\bar{b}\bar{c}+\bar{a}b\bar{c}+\bar{a}\bar{b}c  \\
&= abc+(a+b+c)(\bar{a}\bar{b}+\bar{b}\bar{c}+\bar{a}\bar{c})   \\
&= abc + (a+b+c)\bar{M}_{3}  \\
&= abc + (a+b+c+ab+ac+bc)\bar{M}_{3}   \\
&= abc+a\bar{M}_{3}\bar{M}_{3}+b\bar{M}_{3}\bar{M}_{3}+c\bar{M}_{3}\bar{M}_{3}
\\
&~~~+ab\bar{M_{3}}+ac\bar{M_{3}}+bc\bar{M_{3}} , 
\end{split}\end{equation}
where $M_{3}$ represents the 3-input majority logic $M_{3}(a,b,c)$. The final expression indicates a 5-input majority logic of $M_{5}(a,b,c,\bar{M}_{3},\bar{M}_{3})$.
\begin{figure}
\includegraphics[width=5.0cm]{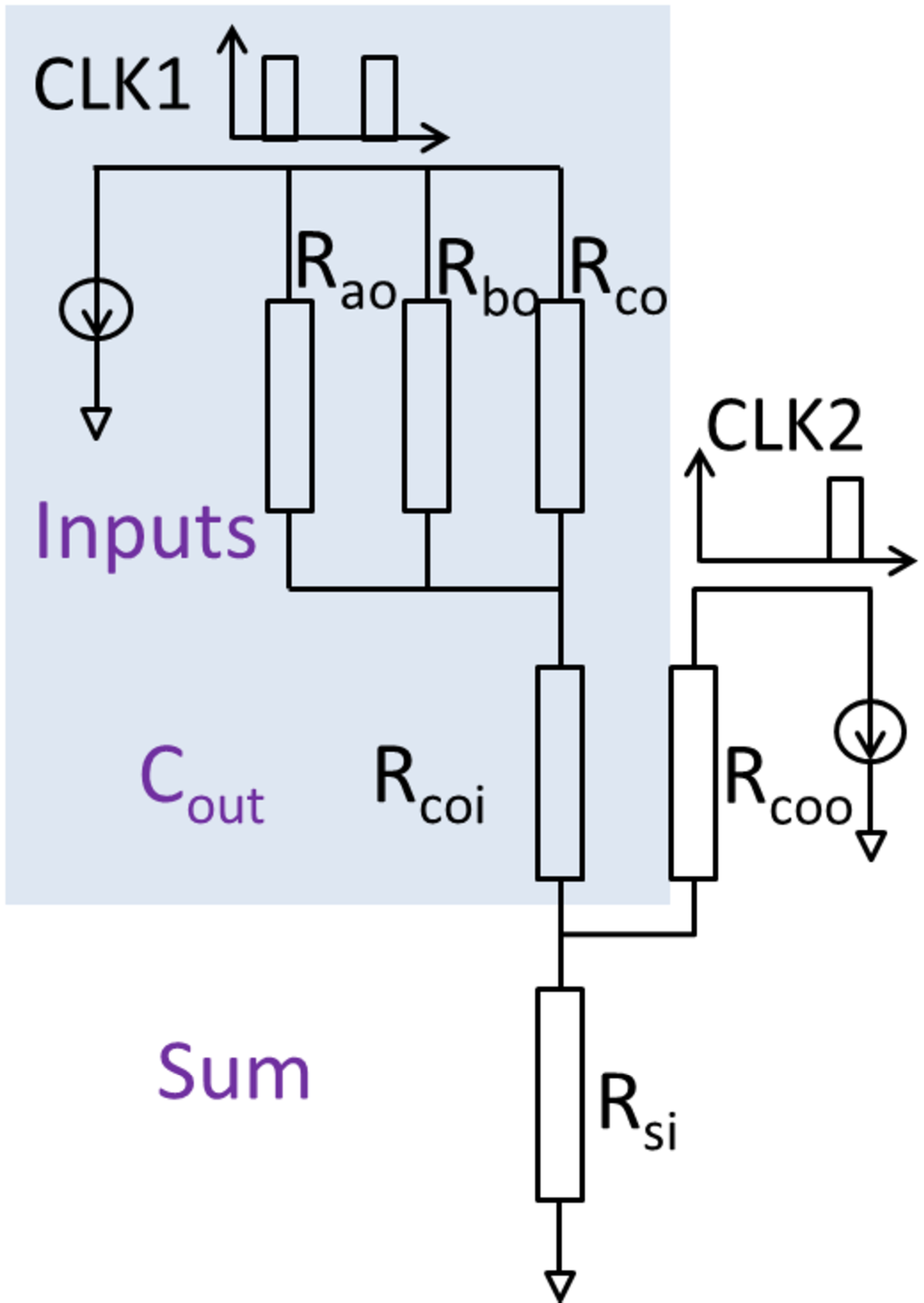}
\caption{
Equivalent circuit of the 1-bit adder shown in Fig.~5(a) of the main paper. The shaded part indicates the 3-input majority gate at the first stage.  Each cell is represented by its resistance that is driven by the first pulse of CLK1. CLK2 is synchronized with the second pulse of CLK1 to drive the second stage at which the entire circuit is active, realizing a 5-input majority logic. The spin dependent conductance is treated separately at each resistor, while the total current is used for the circuit simulation at each node.}
\label{fig_adderc}
\end{figure}

A similar analysis has been applied to decompose the adder into 3-input majority gates based on quantum cellular automata that only enables three inputs.~\cite{Zhang2004} In contrast, the device proposed here relies on the ensemble of all injected electrons and there is no limit in the number of inputs. By using the input signal ($a$,$b$,$c$) more than once in addition to the intermediate output $c_\mathrm{out}$, the add operation can be realized by a minimum of 5 elemental cells [Fig.~5(a)] in the main paper); a comparable type of adder implementation has previously been used in the metallic spin device proposals as well.~\cite{Augustine2011}  The equivalent circuit model of this particular layout is shown in Fig.~\ref{fig_adderc}.

As an estimate for the total current to ensure the intended functionality, the worst case is identified to have one of the inputs opposite to the other two, e.g., $a=1$, $b=1$, $c=0$. Assuming that logic state "$1$" corresponds to the magnetization along the $+x$ direction (i.e., high channel conductance for spin $+1/2$ electrons), the total conductance for the spin "$+1/2$" channel is $2G_{H}+G_{L}$ while that for the spin "$-1/2$" channel is $G_{H}+2G_{L}$.  Here $G_{H}$ ($G_{L}$) represents the higher (lower) conductance value (or, equivalently, current) for the preferred (not preferred) spin. This results in the total polarization of $\frac{G_{H}-G_{L}}{3(G_{H}+G_{L})}\approx\frac{1}{3}$ to write the $\mathrm{C_{out}}$ cell at the first clock (CLK1). Similarly, the electron polarization to write the $S$ cell at the second stage (CLK2), which is effectively a 5-input majority gate, is only about $\frac{1}{5}$. As a result, the current necessary to achieve the set-level of the spin signal strength (i.e., spin polarized current) must increase proportionally with the number of inputs. Given the parallel connections of these inputs, the power consumption would also increase approximately linearly.

\section{Circuit simulation}

\begin{figure}
\includegraphics[width=8.0cm]{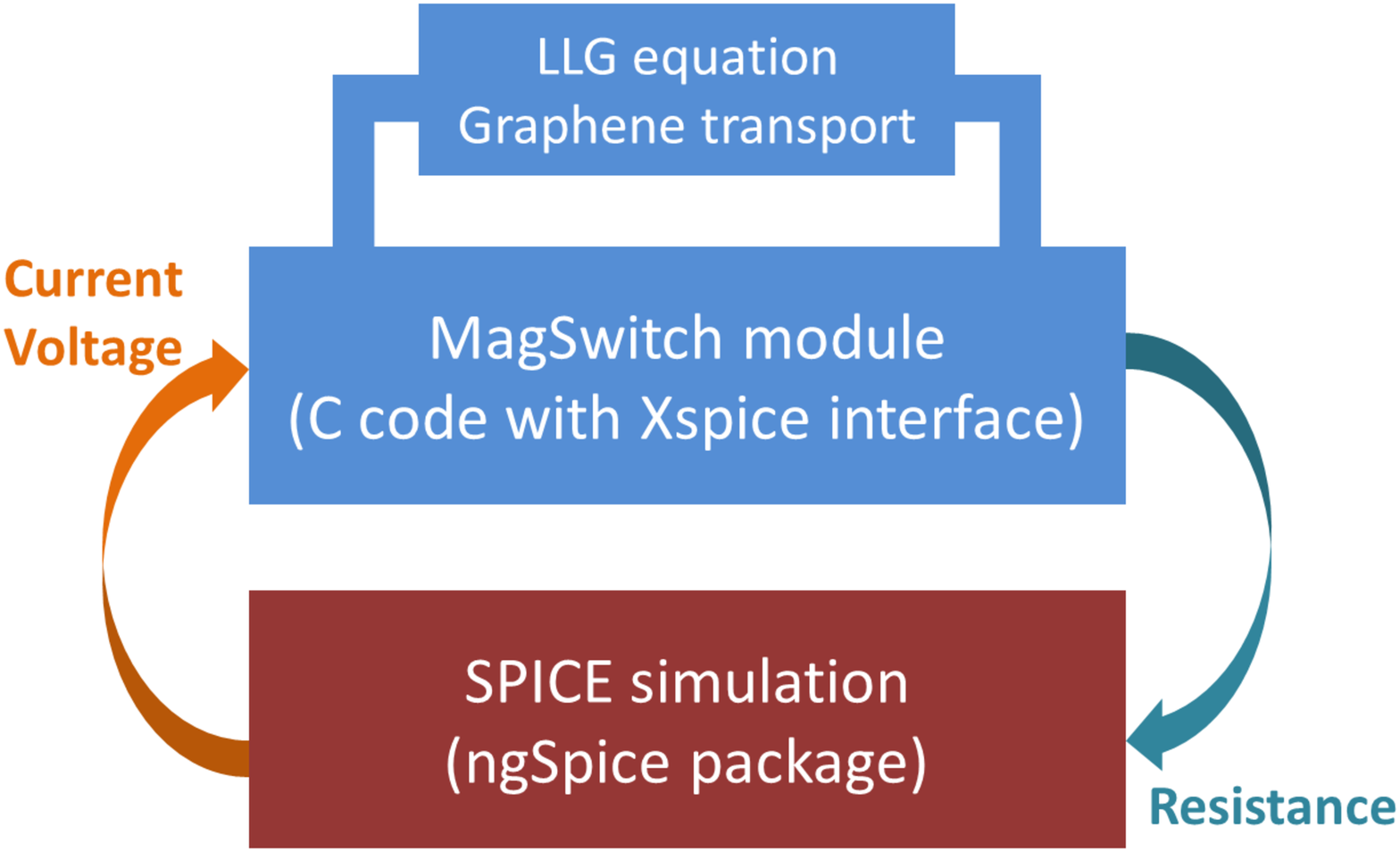}
\caption{
Schematic illustration of the simulation frame work.}
\label{fig_simu}
\end{figure}

The extracted equivalent circuit is implemented as a XSPICE user defined model in the ngSpice circuit simulator as illustrated in Fig.~\ref{fig_simu}. The model contains a LLG solver to simulate the magnetization dynamics. In each step of transient simulation, the magnetization state is updated according to the LLG solver and the result is used to determine the resistance values. These values are then adopted in the circuit simulation.  Thus, our procedure realizes a circuit-device co-simulation, as adopted in other numerical studies of spin logic circuits,~\cite{Augustine2011,Sharad2012} plus a real-time LLG solver. The simulation set-up is applied to verify the 1-bit adder operation as show in Fig.~5(b) (the main paper).

\section{Spin transfer via carrier mediated exchange coupling}
The Hamiltonian for the graphene electron can be expressed as:
\begin{equation}
H=\hbar v_{F}\mathbf{k}\cdot\mathbf{\sigma}+2G_{0}\mathbf{m}(x)\cdot\mathbf{S},  \label{h12}
\end{equation}
where $\mathbf{k}$ is the in-plane electron wave vector [$= -i(\partial /\partial x,\partial /\partial y,0)$], $\mathbf{\sigma}$ the Pauli matrix vector defined on the graphene sublattice states, $G_{0}$ the exchange constant (as introduced previously), and  $\mathbf{S}$ the electron spin operator. $\mathbf{m}(x)$ takes the value $\mathbf{m}_i$ of the corresponding region with the zero of the $x$ axis set to the boundary between $\mathbf{M}_{1}$ and $\mathrm{\mathbf{M}_{C}}$: i.e., $\mathbf{m}(-L < x <0)= \mathbf{m}_1$, $\mathbf{m}(0 <x<L_{\mathrm{C}})=\mathrm{\mathbf{m}_C}$, $\mathbf{m}(L_{\mathrm{C}}<x<L_{\mathrm{C}}+L)=\mathbf{m}_2$, where $L$ and $L_{\mathrm{C}}$ denote the length of the input/output and control magnets respectively; $\mathrm{\mathbf{m}_{1,2,C}}$ are their normalized magnetization.  The narrow clearances between the magnets (i.e., the ungated regions) are ignored for simplicity  as they are assumed to be smaller than the screening length which is typically several tens of nanometers in graphene\cite{Giannazzo2009}. The negligible spin-orbital interaction in graphene permits the eigenenergy to separate the contributions as $E_{k,\zeta }= E_{k} \pm \Delta_{k}(\mathbf{m}_1,\mathbf{m}_2,\mathrm{\mathbf{m}_C})$, where $2\Delta _{k}(\mathbf{m}_1,\mathbf{m}_2,\mathrm{\mathbf{m}_C})$ is the spin splitting of the orbital state $k$ by the proximity exchange interaction. It is a function of $\mathbf{m}_1$, $\mathbf{m}_1$ and control magnet $\mathrm{\mathbf{m}_C}$. This is the crucial parameter that determines Eq.~(3) in the main paper. If we pin $\mathrm{\mathbf{m}_1}$ and $\mathrm{\mathbf{m}_C}$ ($\mathrm{\mathbf{m}_1}\perp\mathrm{\mathbf{m}_C}$), it becomes a function of  $\mathrm{\mathbf{m}_2}$ alone, which eases the discussion while capturing all the characters. Note that we take the NOT gate as the example here and the treatment of a COPY gate follows accordingly. 

The estimation of spin splitting $2\Delta _{k}$ requires solving the eigenvalue problem of the Hamiltonian [Eq.~(\ref{h12})]. 
In each of the magnet capped regions ($Q=1,2,3$), the eigenstates can be expressed in the basis of $(+,1/2)$, $(+,-1/2)$, $(-,1/2)$, $(-,-1/2)$ as:
\begin{equation}
\Psi _{Q}=\left(
\begin{array}{c}
\psi _{Q,1} \\
\psi _{Q,2} \\
\psi _{Q,3} \\
\psi _{Q,4} %
\end{array}%
\right), \label{Psi}
\end{equation}%
where $\pm$ represent the two lattice sites and $\pm 1/2$ the electron spin.
By writing the general solution as a linear combination of the eigenstates, i.e.,
\begin{equation}
\Phi _{Q}=\sum_{j=1,2,3,4}c_{Q,j}\psi _{Q,j}e^{i\mathbf{k}_{j}\cdot\mathbf{r} }, \label{Phi}
\end{equation}%
the problem is reduced to finding twelve coefficients $c_{Q,j}$ that are determined by the specific boundary conditions. Firstly, the wave function must satisfy the continuity condition at the two interfaces ($x=0$ and $x= L_{\mathrm{C}}$). Secondly,  the channel edges are treated as gapped graphene
with additional term $E_{g}S_{Z}$ in Eq.~(\ref{h12}) in the extreme limit $%
E_{g}\rightarrow \infty $. Then the calculations lead to:
\begin{eqnarray}
\psi _{1,1} &=&-i\psi _{1,3};\psi _{1,2}=-i\psi _{1,4},  \label{bc1} \\
\psi _{3,1} &=&i\psi _{3,3};\psi _{3,2}=i\psi _{3,4}.  \label{bc2}
\end{eqnarray}%

For non-trivial solutions to exist, the following equation needs to be satisfied for the case $\mathbf{m}_{2}=\mathbf{m}_{1}$:
\begin{equation}
\begin{split}
2\cos&{[2\frac{E}{\hbar v_{F}}(2L+L_{\mathrm{C}})]}= 1-\cos{(2\frac{G_0}{\hbar v_F}L_{\mathrm{C}})} 
\\
&-\cos{(4\frac{G_0}{\hbar v_F}L)} - \frac{1}{2}\lbrace\cos{[2\frac{G_0}{\hbar v_{F}}(2L-L_{\mathrm{C}})]}
\\
&+\cos{[2\frac{G_0}{\hbar v_{F}}(2L+L_{\mathrm{C}})]}\rbrace.
\end{split}
\label{ge}
\end{equation}%
In the simplest scenario of zero spin splitting [i.e., $G_{0}=0$], Eq.~(\ref{ge}) reduces to
\begin{equation}
\cos{[\frac{E}{\hbar v_{F}}(2L+L_{\mathrm{C}})]}=0  \label{g0}
\end{equation}%
that recovers the well-known result for graphene electron confinement:~\cite{Akhmerov2008,Tworzydo2006}
\begin{equation}
E_{n,\zeta }=(n+\frac{1}{2})\frac{\pi\hbar v_{F} }{2L+L_{\mathrm{C}}},~~~~ n=0,\pm 1,... \label{eg0}
\end{equation}%
Note that each $E_{n,\zeta }$ in Eq.~(\ref{eg0}) is doubly degenerate in the spin index $\zeta$ (=$\pm \frac{1}{2}$). When the input and target magnets are coupled antiferromatically (i.e., $G_{0}=\hbar v_{F}\pi /2L_{\mathrm{C}}$), on the other hand, Eq.~(\ref{ge}) results in
\begin{equation}
\sin{[\frac{E}{\hbar v_{F}}(2L+L_{\mathrm{C}})]}=0.  \label{ef}
\end{equation}%
This generates the spectrum
\begin{equation}
\begin{split}
E_{n,\zeta }&=(n+\frac{1}{2})\frac{\pi \hbar v_{F}}{2L+L_{\mathrm{C}}}+\zeta \frac{\pi \hbar v_{F}}{2L+L_{\mathrm{C}}}
\\
&=(n+\frac{1}{2})\frac{\pi \hbar v_{F}}{2L+L_{\mathrm{C}}}+\zeta \frac{2L_{\mathrm{C}}G_{0} }{2L+L_{\mathrm{C}}}
\end{split}
\label{enx}
\end{equation}
Thus, the spin-splitting energy is
\begin{equation}
2\Delta_k(\mathbf{m}_2=\mathbf{m}_1) =2G_0\frac{\epsilon }{2+\epsilon }.  \label{ssx}
\end{equation}%
where $\epsilon =L_{\mathrm{C}}/L$ (normally $\epsilon<2$ for practical devices). 
Note that the no wavevector dependence is observed in the final result.

A similar procedure can be applied for the case of $\mathbf{m}_2=-\mathbf{m}_1$ with $G_{0}=\hbar v_{F}\pi /2L_{\mathrm{C}}$. Straightforward algebra leads to the energy spectrum in the form%
\begin{equation}
\cos{[\frac{2E}{\hbar v_F}(2L+L_{\mathrm{C}})]}=\cos{(4\frac{G_0}{\hbar v_F}L)} ,  \label{emx}
\end{equation}
\begin{equation}
E_{n,\zeta }=n\frac{\pi \hbar v_F}{2L+L_{\mathrm{C}}}+\zeta \frac{4LG_0}{2L+L_{\mathrm{C}}}.  \label{enm}
\end{equation}%
The corresponding spin splitting is then
\begin{equation}
2\Delta_k(\mathbf{m}_2=-\mathbf{m}_1) =2G_0\frac{2}{2+\epsilon }.  \label{ssm}
\end{equation}

Following this method, the other cases for the $\mathbf{m}_{2}$ configurations directed along the $\mathbf{m}_{C}$ axis can also be calculated.
\begin{eqnarray}
\Delta _{k}(\mathbf{m}_2=\mathbf{m}_C) &=&\frac{1+\epsilon }{2+\epsilon }G_{0};  \label{yp}
\\
\Delta _{k}(\mathbf{m}_2=-\mathbf{m}_C) &=&\frac{1}{2+\epsilon }G_{0};  \label{ym}
\end{eqnarray}
A further analysis indicates that $\mathbf{m}_{2}= -\mathbf{m}_1$ results in the lowest thermodynamic potential for practical devices and is indeed the preferred state of the target magnet.
It is worth noting that these results on energy splitting are expressed in terms of the ratio of the control magnet length over that of the unit cells, which can also be obtained by taking the length as normalization factors for unrestricted graphene layer. Actually in this structure, the transverse quantum confinement is negligible for practical device size (e.g., 100 nm) as the thermal energy  $k_{B}T$ is well above quantization step $\Delta E=\pi /(2L+L_{\mathrm{C}})$ at room temperature.

\begin{figure}
\includegraphics[angle=0, scale=0.5, viewport = 180 85 575 475]{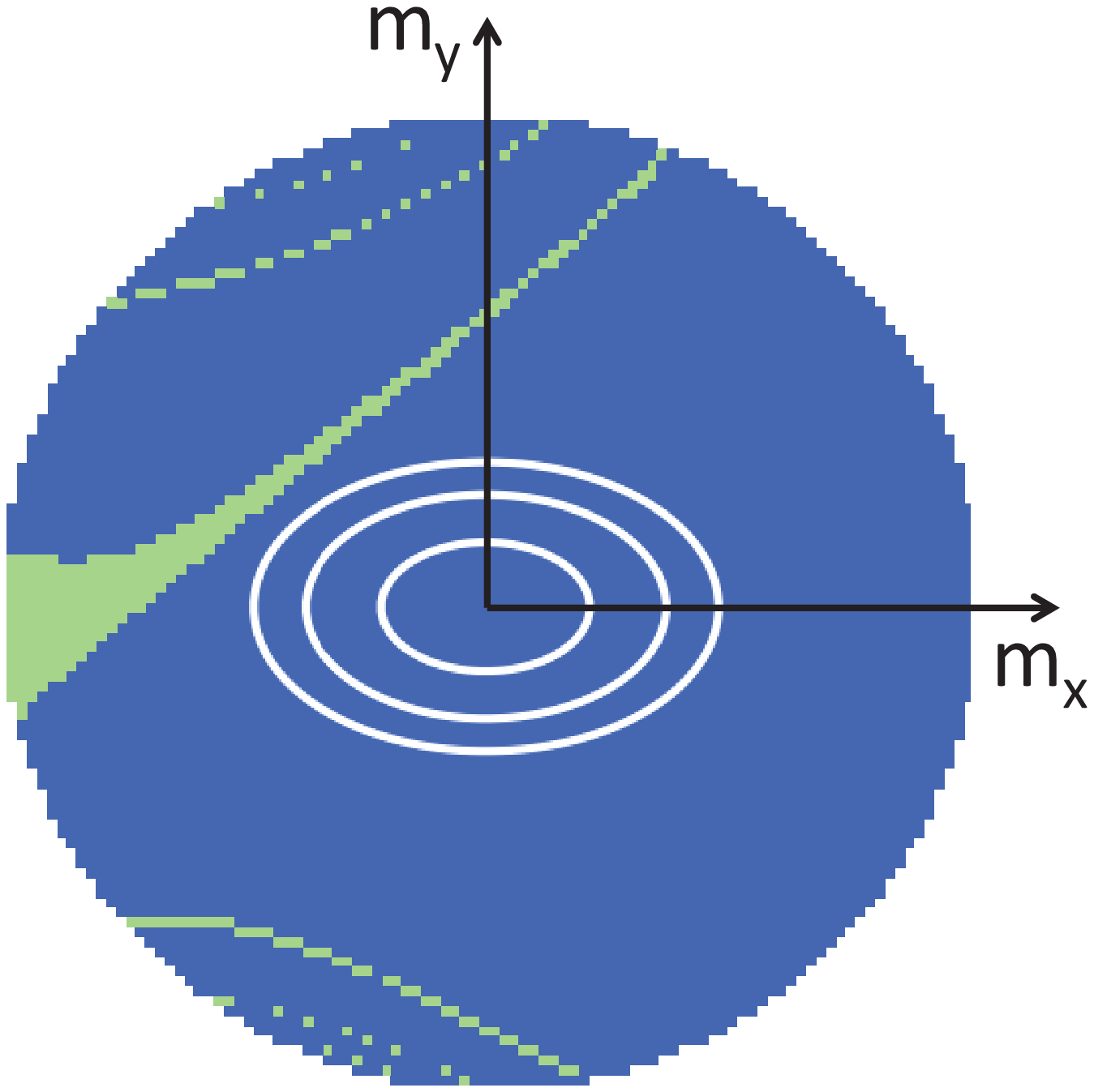}
\caption{
Topology of switching success/failure with the effective field of $H_{x}^{\rm eff}=1500$~Oe , $H_{y}^{\rm eff}=-280$~Oe. The null-state magnetization that relaxes to $m_{x}=1$ is shown in blue (success), while that led to $m_{x}= -1$ is marked in green (failure). The ellipses indicate the energy contours of 2$k_{B}T$, 4$k_{B}T$, and 10$k_{B}T$ from the neutral state.  The magnet has a dimension of $60\times 60\times 2$~nm$^3$ with a hard axis anisotropy $K_{y}=40~\mathrm{fJ/\mu m}$. }
\label{fig_mapv}
\end{figure}

To access the reliability of this approach, since the magnetic switch follows the same dynamics as that characterized  for the current driven case, the critical field strength for a reliable operation can be found by referring to the critical signal current density.  Adopting $J = 0.6~\mu$A/nm, the corresponding effective field gives around $1250$~Oe.   This is further verified by simulated switching result with $H_{x}=1500$~Oe, $H_{y}=-280$~Oe as shown in Fig.~\ref{fig_mapv} as a function of the neutral state magnetic configuration in the target cell. The estimated error rate is less than $10^{-7}$.

\section{Dynamics of magnetization switch via Bennett clocking}
In a structure where a topological insulator (TI) is in contact with a thin magnet, the carrier-ion exchange interaction at the interface can be described by introducing an effective out-of-plane magnetic anisotropy in the magnet as discussed in detail in Ref.~\onlinecite{Semenov2012}.  Similarly, the influence of the signal pulse for a deterministic $180^{\circ}$ switch may be accounted for by an additional torque ($\mathbf{T}$). Thus, the magnetization dynamics in our treatment is modeled phenomenologically by extending the Landau-Lifschitz-Gilbert (LLG) equation as:
\begin{equation}
\frac{\partial{\mathbf{m}}}{\partial{t}}=-\gamma\mathbf{m}\times\mathbf{H}_\mathrm{ eff}+\alpha\mathbf{m}\times\frac{\partial{\mathbf{m}}}{\partial{t}}+\mathbf{T} ,
\label{eq_LLG}
\end{equation}
where $\mathbf{m}$ is the normalized magnetization defined as $\mathbf{m}=\mathbf{M}/|\mathbf{M}|$,  $\gamma$  the gyromagnetic ratio, and $\alpha$ the  Gilbert damping factor. The effective magnetic field $\mathbf{H}_\mathrm{eff}$ accounts for the effects of demagnetization field and other anisotropy terms (including the induced out-of-plane anisotropy discussed above).~\cite{Semenov2012}  The torque is treated as the effect of an exchange field due to the polarized electrons:
\begin{equation}
\mathbf{T}=\frac{\gamma G_0}{\mu_{0}M_{0}L_{z}}\langle\mathbf{\sigma}_{s}\rangle\times\mathbf{m} ,
\label{eq_Tcurr}
\end{equation}
where $G_{0}$ represents the coupling constant for the proximity exchange interaction in energy units, $\mu_{0}$ the permeability constant, $M_{0}$ the saturation magnetization ($M_{0}=|\mathbf{M}|$), $L_{z}$ the thickness of the magnet, and $\mathbf{\sigma}_{s}$ the vector spin operator in the form of Pauli matrices.

More specifically, the electron spin polarization $\langle \mathbf{\sigma}_{s}\rangle$ depends on the channel properties.  In a TI surface channel, where the electron spin is locked to its momentum, the electric current $\mathbf{J}=(J_{x},J_{y},0)$ polarizes electron spin~\cite{Yokoyama2011} as $\langle\mathbf{\sigma}_{s}\rangle=(-\frac{J_{y}}{ev_{F}},\frac{J_{x}}{ev_{F}},0)$.  In other systems like a nonmagnetic metal or graphene, on the other hand, the current itself does not evoke spin polarization; hence, it needs an additional interaction with a magnetic material to polarize the electrons.  When a spin polarized current $\mathbf{J}^{\uparrow}=(J_{x}^{\uparrow},J_{y}^{\uparrow},0)$ is present, the electron polarization in these materials can be defined as $\langle\mathbf{\sigma}_{s}\rangle=(\frac{J_{x}^{\uparrow}}{ev_{F}}, \frac{J_{y}^{\uparrow}}{ev_{F}},0)$.  Here, $\uparrow$ is used simply to symbolize the polarized nature.  As shown, the two cases give comparable expressions and the impact on the dynamics of $\mathbf{M}$ can be handled essentially independent of the origin (i.e., the TI or graphene channel) once the spin polarization on the surface is specified.  In the present analysis, we designate the in-plane easy axis as the $x$ direction for convenience.   It is also assumed that the input signal is applied in such a way to induce the channel polarization along the same coordinate axis (i.e., $\langle\mathbf{\sigma}_{s}\rangle \parallel \pm \hat{{\mathbf x}}$).

We numerically solve Eq.~(\ref{eq_LLG}) using the Runge-Kutta method to characterize the switching dynamics. The magnet is chosen to have a dimension of $60\times 60\times 2$~nm$^{3}$, saturation magnetization $M_{0}=160$~Oe, and the damping factor $\alpha=0.1$. The exchange constant $G_{0}$ is set to $40$~meV .  While this is not a well characterized parameter, it is not without the relevant studies in the literature. A recent first principle calculation\cite{Luo2013} reported a band gap of $54~\mathrm{meV}$ in MnS capped Bi$_2$Se$_3$ (thus, the exchange coupling constant of 27 meV).  Given the low N{\'e}el temperature of antiferromagnetic MnS (about 170 K),~\cite{Jacobson1970}  significantly larger values can be expected for magnetic materials with higher N{\'e}el or Curie temperatures. Indeed, moderate magnetic doping of TI with Cr has shown a gap opening over 120 meV experimentally.~\cite{Kou2012} Moreover, a similar measurement in the closely related Ni-graphene structures resulted in the exchange bias field as large as 2000 Oe (between two thin Ni layers separated by graphene),~\cite{Mandal2012} which corresponds to the proximity coupling constant of around 240 meV. Accordingly, one can reasonably anticipate $G_{0}$ to be in the tens to hundreds of meV at the well prepared TI/magnet or graphene/magnet interfaces.